\documentclass[twocolumn,prd,nofootinbib,aps,floats,floatfix,amsmath,amssymb,secnumarabic]{revtex4} %
\pdfoutput=1
\usepackage{graphicx}
\usepackage[usenames,dvipsnames]{color}
\usepackage{amsmath,amssymb}
\usepackage{slashed}
\usepackage[colorlinks,citecolor=blue]{hyperref}
\def\sfrac#1#2{{\textstyle{#1\over #2}}}
\newcommand{\be}{\begin{equation}}
\newcommand{\ee}{\end{equation}}
\newcommand{\ba}{\begin{array}}
\newcommand{\ea}{\end{array}}
\newcommand{\bea}{\begin{eqnarray}}
\newcommand{\eea}{\end{eqnarray}}
\newcommand{\sss}{\scriptscriptstyle}

\newcommand{\W}{{\sss W}}
\newcommand{\Z}{{\sss Z}}

\renewcommand{\S}{{\sss S}}

\def\sfrac#1#2{{\textstyle{#1\over #2}}}
\newcommand{\lhs}{\lambda_{h\S}}
\newcommand{\Fermi}{Fermi}
\newcommand{\Planck}{Planck}

\begin{document}
\title{Update on scalar singlet dark matter}
\author{James M.\ Cline}
\email{jcline@physics.mcgill.ca}
\affiliation{Department of Physics, McGill University,
3600 Rue University, Montr\'eal, Qu\'ebec, Canada H3A 2T8}
\author{Kimmo Kainulainen}
\email{kimmo.kainulainen@jyu.fi}
\affiliation{Department of Physics, P.O.Box 35 (YFL),
FIN-40014 University of Jyv\"askyl\"a, Finland}
\affiliation{Helsinki Institute of Physics, P.O. Box 64,
FIN-00014 University of Helsinki, Finland}
\author{Pat Scott}
\email{patscott@physics.mcgill.ca}
\affiliation{Department of Physics, McGill University,
3600 Rue University, Montr\'eal, Qu\'ebec, Canada H3A 2T8}
\author{Christoph Weniger}
\email{c.weniger@uva.nl}
\affiliation{GRAPPA Institute, University of Amsterdam,
Science Park 904, 1098 GL Amsterdam, Netherlands}

\begin{abstract}
One of the simplest models of dark matter is that where a scalar singlet field $S$ comprises some or all of the
dark matter, and interacts with the standard model through an
$|H|^2 S^2$ coupling to the Higgs boson.  We update the present limits
on the model from LHC searches for invisible Higgs decays, the thermal relic density of $S$, 
and dark matter searches via indirect and direct detection.  We point out that the currently allowed parameter
space is on the verge of being significantly reduced with the next
generation of experiments.  We discuss the impact of such constraints
on possible applications of scalar singlet dark matter, including
a strong electroweak phase transition, and the question of
vacuum stability of the Higgs potential at high scales.  

\end{abstract}
\maketitle

\section{Introduction}

Scalar singlet dark matter \cite{Silveira:1985rk, McDonald:1993ex, Burgess:2000yq} is an attractive model due to its
simplicity; the essential couplings are just its bare mass term and
a cross-coupling to the standard model (SM) Higgs field, 
\be
    V = \sfrac12 \mu_S^2 S^2 + \sfrac12\lhs S^2|H|^2\;.
\label{spot}
\ee
After electroweak symmetry breaking, the $S$ boson mass receives contributions from both terms, giving
\be
    m_\S = \sqrt{\mu_S^2 + \sfrac12{\lhs v_0^2} }\;,
\ee
where $v_0=246.2$\,GeV is the Higgs VEV.  Phenomenology of this model
has been studied in refs.\ \cite{Davoudiasl:2004be, Ham:2004cf, Patt:2006fw, O'Connell:2006wi, He:2007tt, Profumo:2007wc, Barger:2007im, He:2008qm, Ponton:2008zv, Lerner:2009xg, Farina:2009ez, Bandyopadhyay:2010cc, Barger:2010mc, Guo:2010hq, Espinosa:2011ax, Profumo:2010kp, Djouadi:2012zc, Mambrini:2012ue, Drozd:2011aa, Grzadkowski:2009mj}.

The Higgs cross-term is generically expected to be present because it
is a dimension-4 operator that is not forbidden by any symmetry. 
Apart from the $S$ kinetic term and its quartic self-coupling (which
plays no observable role in phenomenology), the two terms in eq.\
(\ref{spot}) are in fact the only renormalizable terms allowed by 
general symmetry arguments.  Terms cubic or linear in $S$ are excluded
if one demands that $S$ is absolutely stable, and therefore a viable dark matter (DM)
candidate, by imposing the $Z_2$ symmetry $S\to -S$.  In this scenario $S$ is a classic weakly-interacting
massive particle (WIMP); although it is possible to make $S$ a viable, metastable DM candidate 
without the $Z_2$ symmetry, here we focus exclusively on the stable case.

The single $S^2|H|^2$ coupling is however enough to allow for a
contribution to the invisible decay of the Higgs boson,  scattering of
$S$ on nucleons through Higgs exchange, and annihilation of $S$ into
SM particles, leading to indirect detection signatures and an allowed
thermal relic density.  The scalar singlet model with $Z_2$ symmetry is, in essence, the
simplest possible UV-complete theory containing a WIMP.  It is intriguing that natural values of
$\lhs\lesssim 1$ and $m_\S$ below a few TeV\footnote{These upper
limits based on perturbativity in the $\lhs$ coupling are more stringent than the unitarity bounds on the 
annihilation cross-section \cite{Griest:1989wd}.} simultaneously
reproduce the observed DM relic density and predict a cross section
for scattering on nucleons that is not far from the current direct
detection limit.

These aspects have of course been widely studied, with refs.\ 
\cite{Mambrini:2011ik, Low:2011kp, Djouadi:2011aa, Cheung:2012xb} 
providing the most recent  comprehensive analyses.   We believe it is
worthwhile to update the results presented there, for several
reasons.  
\begin{enumerate}
\item Some \cite{Mambrini:2011ik, Low:2011kp} were done before the mass of the Higgs boson was measured by ATLAS and CMS, and the dependence of the results on $m_h$ was shown for only a limited number of Higgs masses.
\item With the exception of ref.~\cite{Cheung:2012xb}, these recent studies were performed prior to the release of updated direct detection constraints by the XENON100 Collaboration \cite{Aprile:2012nq}.
\item The predicted direct detection cross section depends on the Higgs-nucleon coupling.  Recent results from lattice studies \cite{Bali:2011ks, Bali:2012qs,Alvarez-Ruso:2013fza,Oksuzian:2012rzb, Gong:2012nw, Freeman:2012ry,
Engelhardt:2012gd, Junnarkar:2013ac, Young:2013nn, Jung:2012rz, Gong:2013vja} and chiral perturbation theory \cite{Gasser:1990ce, Gasser:1990ap, Borasoy:1996bx, Alarcon:2012kn, Alarcon:2012nr, Pavan:2001wz, Cheng:2012qr} have reduced the theoretical uncertainty in this quantity.  
\item Limits on the invisible width of the Higgs have improved \cite{Belanger:2013xza} since all of the recent studies of this model, reducing the allowed parameter space in the region $m_\S< m_h/2$.
\item The constraints on $\lhs$ from direct detection presented by
refs.\ \cite{Mambrini:2011ik, Djouadi:2011aa, Cheung:2012xb} and from
indirect detection in ref.\ \cite{Cheung:2012xb} were
derived without taking into account the fact that larger values of
$\lhs$ suppress the $S$ relic density, by increasing the annihilation
cross section.  This reduces the overall predicted signal for
scattering on nucleons, and annihilation into SM particles.  Because
of this effect, the dependence on $\lhs$ of the direct and indirect
detection constraints is significantly different than one might have
expected, as noted in ref.\ \cite{Low:2011kp}.  We take the view here that singlet dark matter might provide only a fraction of the total dark matter density, which is a logical possibility. 
\item In some previous studies (e.g. ref.~\cite{Low:2011kp}), the relic
density has not been computed using the full thermal average of the
annihilation cross section.  It is necessary to do so when $m_\S$ is
near $m_h/2$ in order to obtain accurate results, because the integral
over DM velocities is sensitive to the degree of overlap with the
resonance in $\sigma v_{\rm rel}$ at centre-of-mass energy $E_{\sss CM} =
m_h$.  This can change the result by orders of magnitude in comparison to using the threshold approximation.
\item So far ref.\
\cite{Cheung:2012xb} has been the only comprehensive study of scalar singlet DM to 
consider recent indirect detection constraints.  The most important of these are  gamma-ray constraints from \Fermi\ observations of 
dwarf galaxies.  Ref.\ \cite{Cheung:2012xb} implemented these limits in an approximate 
fashion, rescaling published 95\% limits on the cross-sections for 
annihilation into an incomplete set of SM final states, and ignoring 
the $SS\to hh$ channel.  Here we calculate constraints 
self-consistently for the complete set of branching fractions to SM 
final states at every point of the parameter space, adding
 further constraints from the impact of $SS$ annihilation on the CMB,
 and providing projected constraints including the impact of the 
\v{C}erenkov Telescope Array (CTA).
\end{enumerate}

In the following, we outline updated constraints and projections from
the Higgs invisible width (section \ref{higgswid}), the $S$ thermal
relic density (section \ref{relicdens}), indirect detection (section
\ref{id}) and direct detection (section \ref{dd}).  The relevance of
these constraints to some applications of the model is discussed in
section \ref{apps}.  We give conclusions in 
section \ref{summary}.

\begin{figure*}[tb]
\includegraphics[width=\columnwidth]{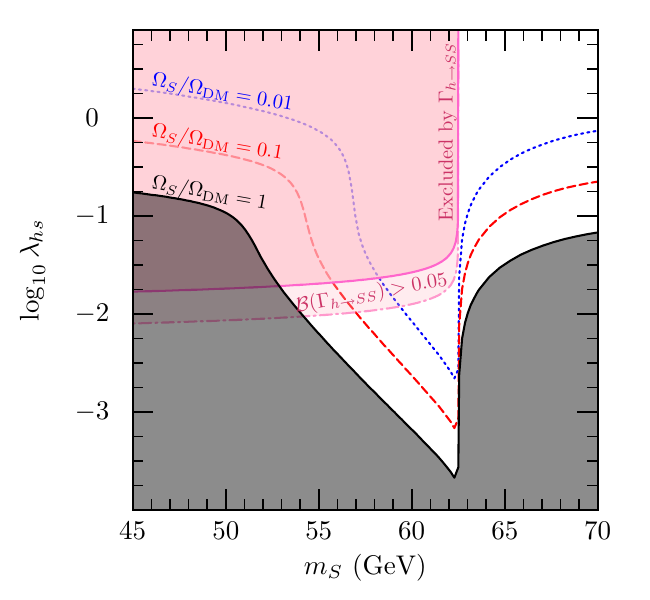}\hspace{0.06\columnwidth}
\includegraphics[width=\columnwidth]{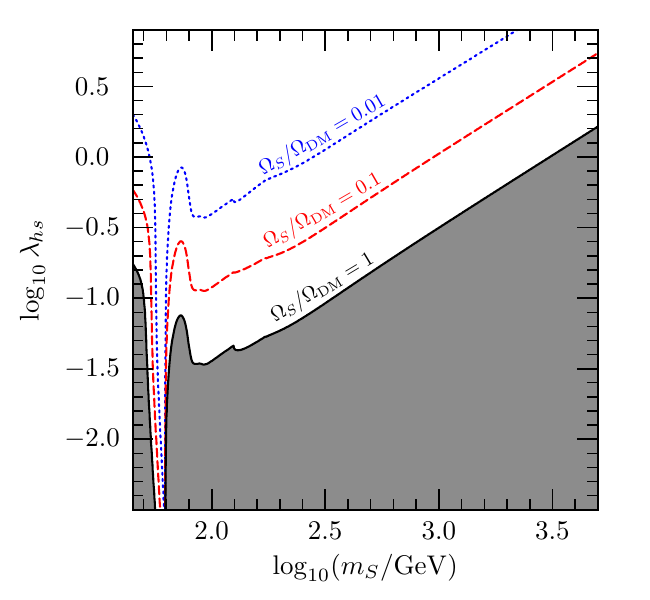}
\caption{Contours of fixed relic density, labelled in terms of their fraction of the
full dark matter density.  Dark-shaded lower regions are ruled out
because they produce more than the observed relic density of dark matter.   
\textit{Left}: a close-up of the mass region $m_\S
\sim m_h/2$, where annihilations are resonantly enhanced.  The region
ruled out by the Higgs invisible width at $2\sigma$ CL is indicated by the
darker-shaded region in the upper left-hand corner.  The projected
$1\sigma$ constraint from 300\,fb$^{-1}$ of luminosity at the 14\,TeV LHC is shown as the
lighter-shaded region, corresponding to a limit of 5\% on the Higgs branching fraction to invisible states \cite{Peskin:2012we}.  \textit{Right}: relic density contours for the full range of $m_\S$. 
}
\label{fig:relden}
\end{figure*}

\section{Higgs invisible width}
\label{higgswid}

For  $m_\S < m_h/2$, the decay $h\to SS$ is kinematically allowed,
and contributes to the invisible width $\Gamma_{\rm inv}$
of the Higgs boson.  The LHC constraints on $\Gamma_{\rm inv}$
continue to improve as the properties of the Higgs boson are 
shown to be increasingly consistent with SM expectations.
Ref.\ \cite{Belanger:2013xza} obtains a limit of 19\% for the
invisible branching fraction at $2\sigma$, based on
a combined fit to all Higgs production and decay channels probed
by ATLAS, CMS and the Tevatron.
 
The contribution to $\Gamma_{\rm inv}$ in the scalar singlet
dark matter model is 
\be
    \Gamma_{\rm inv} = {\lhs^2 v_0^2\over 32\pi m_h}
    \left(1 -4 m_\S^2/m_h^2\right)^{1/2}\;,
\label{Gamma_inv}
\ee
(this corrects a factor of 2 error in eq.\ (3.2) of  
ref.~\cite{Cline:2012hg}).  To compute the branching fraction $\Gamma_{\rm
inv}/(\Gamma_{\rm vis} + \Gamma_{\rm inv})$ we take
the visible contribution to the width to be $\Gamma_{\rm vis} = 
4.07$\,MeV for $m_h=125$\,GeV.

In the left panel of Fig.\ \ref{fig:relden}, we show the limit imposed on the scalar singlet parameter space by the invisible width constraint.  For $m_\S<m_h/2$, couplings larger than $\lhs\sim0.02$--$0.03$ are ruled out.  Here we also show the region of parameter space that is projected to be in more than $1\sigma$ tension with data if no additional Higgs decays are detected at the 14\,TeV LHC after 300\,fb$^{-1}$ of luminosity has been collected.  This corresponds to a limit of 5\% on the invisible Higgs branching fraction \cite{Peskin:2012we}.

\section{Relic density}
\label{relicdens}

The relic density of singlet dark matter is mostly determined by Higgs-mediated
$s$-channel annihilation into SM particles. A
sub-dominant role is played by annihilation into $hh$, via the direct 4-boson
$h^2S^2$ vertex, and $S$ exchange in the $t$ channel. As discussed in ref.\
\cite{Cline:2012hg}, tree-level calculations for $SS$ annihilation into two-body
final states do not give a very accurate approximation close to the
threshold for producing gauge boson pairs, as they
miss the 3- and 4-body final states from virtual boson decays, as well
as QCD corrections for quarks in the final state. However, this can be
overcome by using accurate computations of the full Higgs boson width as a
function of invariant mass $\Gamma(m_h^*)$ from ref.\ \cite{Dittmaier:2011ti},
and
factorizing the cross section for annihilation into all SM particles
except $h$ as
\begin{equation}
 \sigma v_{\rm rel} = \frac{2\lhs^2 v_0^2}{\sqrt{s}}|D_h(s)|^2
\Gamma_h(\sqrt{s}) \,,
\label{sv}
\end{equation}
where
\begin{equation}
|D_h(s)|^2 \equiv \frac{1}{(s-m_h^2)^2 + m_h^2\Gamma_h^2(m_h)} \,.
\label{eq:Dh}
\end{equation}
For $m_\S < m_h/2$, the width in the propagator $D_h(s)$ (but not elsewhere)
must be increased by the invisible contribution due to $h\to SS$. For $m_\S >
m_h$, eq.~(\ref{sv}) must be supplemented by the extra contribution from $SS\to
hh$. The perturbative tree level result for the $SS\to hh$ cross section is given in
appendix~\ref{exact x-sections}.

The tabulation of $\Gamma_h(m_h^*)$ in ref.\ \cite{Dittmaier:2011ti} assumes
that $m_h^*$ is the true Higgs mass, associated with a self-coupling $\lambda =
(m_h^*)^2/2v_0^2$. Here $\lambda \approx 0.13$ is fixed by the true Higgs mass
however, and we find that for $\sqrt{s} \gtrsim 300$\,GeV, we must revert to
perturbative expressions for $\Gamma_h(\sqrt{s})$, or otherwise the Higgs 1-loop
self interactions included in the table of ref.~\cite{Dittmaier:2011ti} begin to overestimate 
the width. Above $m_\S=150$\,GeV we revert to the tree-level
expressions for the decay width, including all SM final states. The expressions we use 
can again be found in appendix~\ref{exact x-sections}. 

To accurately determine the relic density for $m_\S$ in the vicinity
of the resonance at $4m_\S^2\sim m_h$ in eq.\ (\ref{sv}), it is essential to carry
out the
actual thermal average \cite{Gondolo:1990dk}
\be
 \langle\sigma v_{\rm rel}\rangle = \int_{4m_\S^2}^\infty
 {s\sqrt{s-4m_\S^2}
 \,K_1(\sqrt{s}/T)\, \sigma v_{\rm rel} \over 16 Tm_\S^4\,
 K_2^2(m_\S/T)}\,{\rm d}s \,,
\label{thermsv}
\ee
where $K_1$, $K_2$ are modified Bessel functions of the second kind, and to
solve the Boltzmann equation for the relic abundance~\cite{Lee:1977ua}.

\begin{figure}[tb]
\includegraphics[width=\columnwidth]{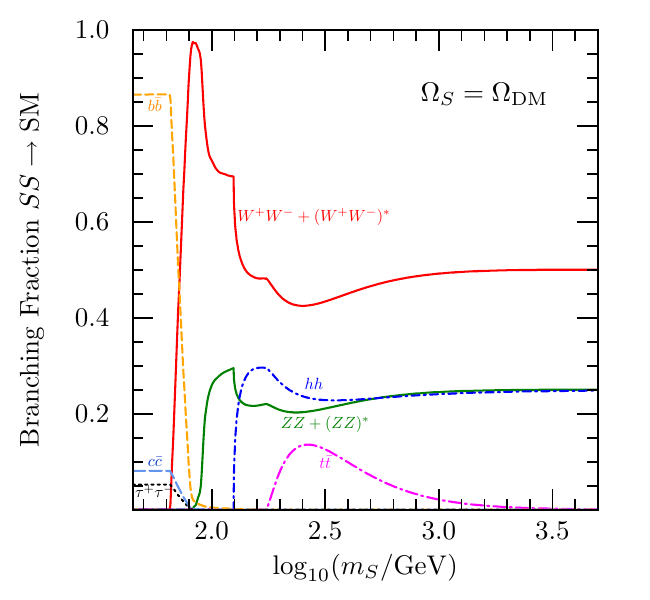}
\caption{Branching fractions for $SS$ to annihilate at threshold into various
SM final states, versus the DM mass. We have chosen $\lhs$ at each dark matter mass such that the $S$ relic density exactly matches the observed value; these $\lhs$ values can be seen along the $\Omega_\S = \Omega_{\rm DM}$ curve in Fig.\ \protect\ref{fig:relden}.} 
\label{figbr}
\end{figure}

\begin{figure*}[tb]
\includegraphics[width=\columnwidth]{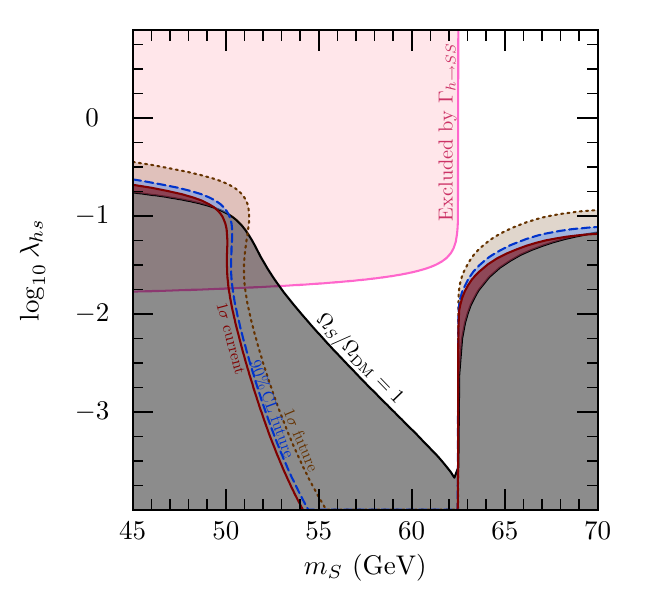}\hspace{0.06\columnwidth}
\includegraphics[width=\columnwidth]{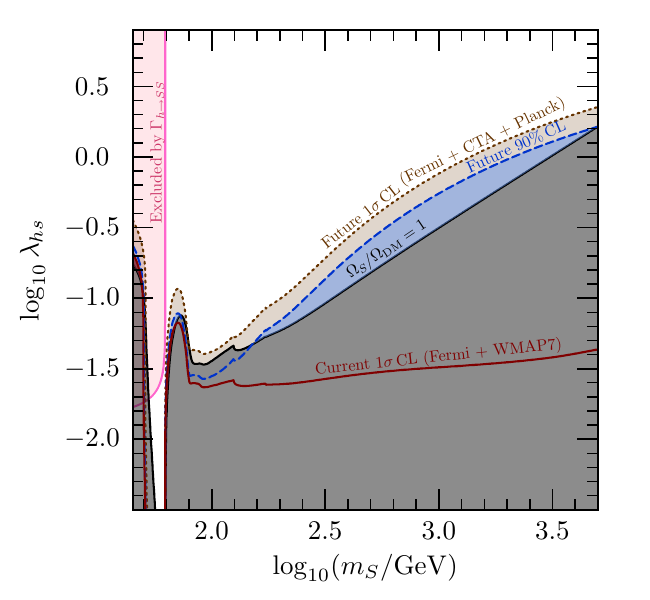}
\caption{Limits on scalar singlet dark matter from indirect searches for dark
matter annihilation. The lowermost shaded region is ruled out because these
models exceed the observed relic density. Regions below the other curves are in
tension with indirect searches, or will be in the future: at more than $1\sigma$
according to current data from \Fermi\ dwarf galaxy observations and WMAP 7-year
CMB data (solid), at $\ge$90\% CL (dashes) and $\ge1\sigma$ CL (dots) with CTA,
\Planck\ polarization data and future \Fermi\ observations. The area ruled out
by the Higgs invisible width at $2\sigma$ CL is indicated by the shaded region in
the upper left-hand corner of both plots. Note that all indirect detection signals are scaled for the thermal relic density of the scalar singlet, regardless of whether that density is greater than or less than the observed density of dark matter. \textit{Left}: a close-up of the
resonant annihilation region. \textit{Right}: the full mass range.}
\label{fig2}
\end{figure*}

\begin{figure}[tb]
\includegraphics[width=\columnwidth]{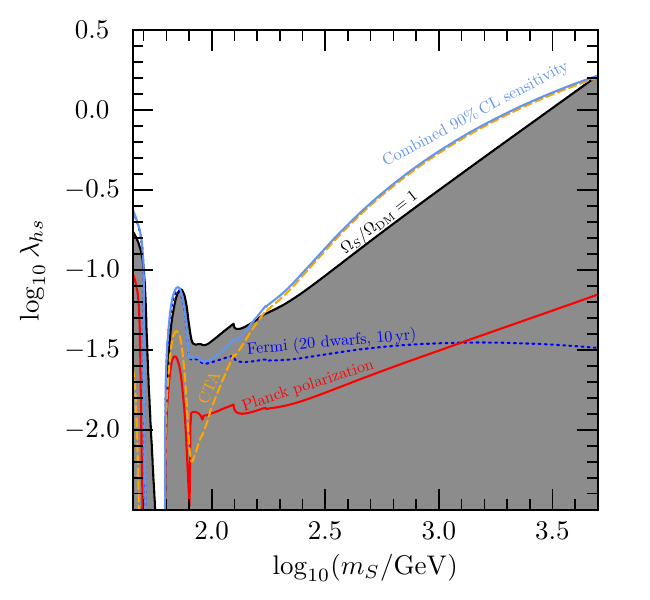}
\caption{Contributions of different searches for dark matter annihilation to the
combined future 90\% CL exclusion curve. The limit from future \Fermi\ searches
for annihilation in dwarf galaxies alone are shown by the dotted line, assuming
10 years of exposure and the discovery of a further 10 southern dwarfs. The
impact of \Planck\ alone, including polarization data, can be seen from the
solid line, and the projected impact of CTA is shown as a dashed line. The
parameter space excluded by the relic density appears once more as a dark shaded
area in the lower part of the plot.}
\label{fig3}
\end{figure}

The common approximation of setting the threshold value of $\sigma v_{\rm rel}$
to the standard value of 1\,pb$\cdot c$ fails badly close to the resonance.
This is because the integral in eq.\ (\ref{thermsv}) can be dominated by the
resonance at $s=m_h^2$ even if $m_\S$ is considerably
below $m_h/2$, possibly increasing $\langle\sigma v_{\rm rel}\rangle$ by orders
of magnitude relative to the threshold value. If $m_\S \gtrsim m_h/2$, the
thermal averaging pushes
$\langle\sigma v_{\rm rel}\rangle$ to lower values relative to the naive
approximation. We compute $\langle\sigma v_{\rm rel}\rangle$ as a function of
temperature and solve the equation for the number density of thermal relic WIMPs numerically,\footnote{We henceforth refer to this as the `Lee-Weinberg equation' with reference to ref.\ \cite{Lee:1977ua}, but note that it has also appeared earlier, e.g. in ref.\ \cite{1966SvPhU...8..702Z}.} using both a full numerical integration and a very accurate approximation described in appendix \ref{appA}. The two methods agree to within
less than~1\%.

The resulting contours of constant relic density are shown in the
plane of $m_\S$ and the coupling $\lhs$ in \mbox{Fig.\ \ref{fig:relden}}. We
display them both over the entire likely range
of dark matter mass values ($45\,{\rm\,GeV} \le m_\S \le 5\,{\rm TeV}$), and in
the region $m_\S\sim m_h/2$ where
annihilation is resonantly enhanced. Constraints from the
Higgs invisible width are also plotted in the low-mass region. Below $m_h/2$,
the two constraints combine to rule out all but
a small triangle in the $m_\S$--$\lhs$ plane,
including masses in the range $52.5-62.5$\,GeV. In the region above
$m_h/2$, the relic density constrains the coupling as a function of mass
in a way that can be approximately fit by the dependence 
$\log_{10}\lhs > -3.63 + 1.04 \log_{10} (m_\S/{\rm\,GeV})$.
We plot up to $\lhs\sim8$, which is at the (generous) upper limit of where the
theory can be expected to remain perturbative.

\section{Indirect detection}
\label{id}

Annihilation of scalar singlet DM into SM particles offers similar
opportunities for indirect detection as with other WIMP DM candidates
\cite{Salati:2010rc,
Porter:2011nv,Yaguna:2008hd,Goudelis:2009zz,Arina:2010rb}.
The strongest current limits come from gamma-ray searches for annihilation in
dwarf spheroidal galaxies \cite{Essig:2009jx, Scott:2009jn, Abdo:2010ex,
Essig:2010em, GeringerSameth:2011iw, Ackermann:2011wa, Tsai:2012cs} (for a
recent general review
see ref.~\cite{Bringmann:2012ez}) and impacts of DM
annihilation at $z\sim600$ on the angular power spectrum of the cosmic
microwave background (CMB) \cite{Padmanabhan:2005es, Natarajan:2009bm,
Galli:2009zc, Slatyer:2009yq, Galli:2011rz, Finkbeiner:2011dx, Slatyer:2012yq,
Cline:2013fm, Weniger:2013hja}.  At large WIMP masses, it is expected
\cite{CTA,CTA_Snowmass} that CTA will provide strong constraints.

We calculate limits on the scalar singlet parameter space implied by indirect detection using a combined likelihood function
\begin{align}
\ln \mathcal{L}_{\rm total}(m_\S,\lhs) &= \ln \mathcal{L}_{\rm CMB}(m_\S,\lhs) \nonumber \\
                                       &+ \ln \mathcal{L}_{\rm dwarfs}(m_\S,\lhs) \\
                                       &+ \ln \mathcal{L}_{\rm CTA}(m_\S,\lhs).\nonumber 
\label{liketot}
\end{align}
In general $\mathcal{L}_{\rm total}$ includes components from all three
indirect searches, but we only include CTA when discussing projected limits.
All three likelihood functions depend in a direct sense upon $m_\S$, but only
indirectly upon $\lhs$, via the zero-velocity annihilation
cross-section $\langle\sigma v_{\rm rel}\rangle_0$, the branching
fractions $r_i$ to the $i$th SM annihilation channel, and the total relic density.

We scale all indirect signals for the appropriate relic density for each combination of $m_\S$ and $\lhs$ self-consistently, suppressing signals where $S$ constitutes only a fraction of the total dark matter.  Where the thermal relic density of $S$ is actually \textit{larger} than the observed dark matter relic density, we simply rescale signals in exactly the same way, increasing the expected signals.  We choose to do this rather than fix the relic density to the observed value in this region for the sake of simplicity and illustration; this region is robustly excluded anyway by the relic density constraint, and the thermal abundance could only be reduced to the observed value if some additional non-thermal effects were added to the scalar singlet theory, which would not be in the spirit of our analysis here.

We calculate $\langle\sigma v_{\rm rel}\rangle_0$ including all allowed two-body SM final
states as per eqs.\ (\ref{sv}) and (\ref{svhh}) for $m_\S\le m_t$ or
eqs.~(\ref{svhh}) and (\ref{GGx-section}-\ref{ffx-section}) for $m_\S>m_t$.
To estimate $r_i$, we calculate $\langle\sigma v_{\rm rel}\rangle_{0,i}$ for
annihilation into a given channel $i$ using these cross-sections\footnote{For determining branching fractions we simply use the tree-level versions; the QCD 1-loop correction has minimal impact above $\sim$$70$\,GeV, and below this the exact partitioning into $b$, $c$ and $\tau$ has only a small effect on integrated gamma-ray yields, so modifies the overall limits from indirect detection only very slightly.}  with the
zero-velocity replacement $\sqrt{s}\rightarrow 2m_\S$, and take $r_i =
\langle\sigma v_{\rm rel}\rangle_{0,i} / \langle\sigma v_{\rm rel}\rangle_0$.
For $m_\S$ just below $m_\W$ and $m_\Z$, where $\langle\sigma v_{\rm rel}\rangle_0$ comes from the factorization approximation, we assign any remaining branching fraction to 3- and 4-body final states arising from annihilation into virtual gauge bosons corresponding to the next most massive threshold, i.e. $(W^+W^-)^*$ for $m_\S<m_\W$ and $(ZZ)^*$ for $m_\W\le m_\S<m_\Z$.

The final yields of photons and electrons from annihilation into each SM final state that we use for CMB limits come from the PPPC4DMID \cite{Cirelli:2010xx}.  The gamma-ray yields we use for \Fermi\ and CTA calculations are from DarkSUSY \cite{Gondolo:2004sc}, which we supplement with the photon yield for the $hh$ annihilation channel from PPPC4DMID.\footnote{For consistency with other channels, we use the $hh$ gamma-ray yields from PPPC4DMID uncorrected for electroweak bremsstrahlung, as none of the DarkSUSY yields take this into account; for all values of $m_\S$ we consider here, the impact of electroweak corrections on the yield from the $hh$ channel is less than 10\%.}  For channels in common, we find good agreement between the gamma yields of PPPC4DMID and DarkSUSY.  

Yields from the 3- and 4-body final states initiated by virtual gauge bosons are also required.  As these are not already available, for \Fermi\ and CTA we estimate the photon yields by analytically extending those of the $WW$ and $ZZ$ channels below threshold.  This is feasible because the integrated photon multiplicity per annihilation in the energy windows considered in each analysis is very close to linear with $m_\S$.  We therefore fit a straight line to this multiplicity over a few GeV above threshold in each case, and use it to extrapolate a small way below threshold ($<10$\,GeV), in the region where the emission of virtual gauge bosons is significant.  This is an extremely good approximation for \Fermi\, and reasonable for CTA also, although not as good as for Fermi due to the energy-dependence of the CTA effective area in this region.  If anything the approximation is marginally optimistic for \Fermi\ (in that the actual yield curve is ever so slightly concave down), whereas for CTA it is conservative (as the true yield curve is slightly concave up).  We do not perform this exercise for CMB limits, as the actual limits near the $W$ and $Z$ thresholds are strongly dominated by \Fermi\ anyway, and it would be more cumbersome to incorporate this into the CMB analysis; we hence assume that 3- and 4-body final states do not contribute anything to CMB limits, which gives a conservative limit in this region.

To show the relative importance of the various final states as a function of $m_\S$, we plot their branching fractions in
Fig.\ \ref{figbr}, along the line in $\{m_\S,\lhs\}$-space where $S$ constitutes the entire observed relic density.  Here we combine the branching fractions of on-shell and off-shell gauge bosons.

\subsection{CMB likelihood}

We take the CMB likelihood function $\mathcal{L}_{\rm CMB}$ directly from the
results presented for annihilation in ref.~\cite{Cline:2013fm} (which were partially based on earlier results in refs.\ \cite{Finkbeiner:2011dx,Slatyer:2012yq}), using tables of the
effective fraction $f_{\rm eff}$ of the DM rest mass injected as additional
energy into the primordial gas.  We interpolate $f_{\rm eff}$ linearly in
$\log m_\S$, then use the calculated values of $r_i$ and $\langle\sigma
v\rangle_0$ for each combination of $m_\S$ and $\lhs$ to obtain the final
likelihood.  We extend the $f_{\rm eff}$ tables of ref.~\cite{Cline:2013fm} in
order to accommodate $S$ masses up to 5\,TeV (see appendix \ref{cmbappendix}
for high-mass $f_{\rm eff}$ data).  For calculating current constraints, we
employ the WMAP 7-year likelihood function~\cite{Komatsu:2010fb}.  For projected constraints we use the \Planck\ predictions, which assume polarization data to be available.  Note that although first \Planck\ TT power spectrum results are available, including limits on DM annihilation \cite{Ade:2013zuv}, these are weaker than projected \Planck\ sensitivities when polarization data is included, and existing WMAP limits.  A factor of a few better constraints than the WMAP7 ones we use are available from WMAP9+SPT+ACT data \cite{Lopez-Honorez:2013cua}, but this improvement will be mostly nullified by a similar degradation in the limits due to corrections to the results of refs.\ \cite{Finkbeiner:2011dx,Slatyer:2012yq}, as discussed in ref.\ \cite{Galli:2013dna}.

\subsection{\Fermi\ dwarf likelihood}
The non-observation of gamma-ray emission from dwarf spheroidal galaxies
by \Fermi\ can be used to put strong constraints on the
annihilation cross-section of dark matter
particles~\cite{GeringerSameth:2011iw, Ackermann:2011wa, Tsai:2012cs}. We
calculate the corresponding \Fermi\ dwarf likelihood function
$\mathcal{L}_{\rm dwarfs}$ based on the results
from ref.~\cite{GeringerSameth:2011iw}, where limits
on the integrated dark matter signal flux with energies from 1 to 100\,GeV
were presented. An alternative treatment with a finer energy binning can be
found in ref.~\cite{Tsai:2012cs}.

From a region $\Delta\Omega$ towards a dwarf spheroidal, one expects a
differential flux of dark matter signal photons that is given by
\begin{equation}
  \frac{{\rm d}\phi}{{\rm d}E} = \frac{\langle \sigma v_{\rm rel}\rangle}{8\pi m_S^2}
    \frac{{\rm d}N_\gamma}{{\rm d}E}
  \underbrace{\int_{\Delta\Omega}{\rm d}\Omega \int_{\rm l.o.s.} \!\!\!{\rm d}s \rho^2}_{\equiv
  J}\;.
\end{equation}
Here, ${\rm d}N_\gamma/{\rm d}E$ denotes the energy distribution of photons produced per
annihilation, and $\int {\rm d}s$ is a line-of-sight integral. The dwarf spheroidals
mainly differ in their dark matter density distribution $\rho$ and
their distance from the Sun, such that the $J$ factor has to be determined
for each dwarf individually. On the other hand, the prefactor is universal.

In~ref.~\cite{GeringerSameth:2011iw}, the authors analyzed the gamma-ray flux from
seven dwarf spheroidals. They determined the probability mass function of the
background events in their signal regions empirically by sub-sampling nearby
regions, and found good agreement with Poisson noise. The $J$ factors of the
individual dwarfs were
adopted from ref.~\cite{Ackermann:2011wa}, and used to define optimized combined
confidence belts that weigh the contribution from each dwarf according to the
probability that observed events belong to the background. This procedure
leads to a combined upper limit on the quantity $\Phi_{\rm PP} \equiv
J^{-1}\int_1^{100\rm \,GeV} {\rm d}E\,{\rm d}\phi/{\rm d}E$. At $95\%$CL, it reads $\Phi_{\rm PP}
\leq 5.0^{+4.3}_{-4.5}\times 10^{-30}\,\rm cm^3\,s^{-1}\,GeV^{-2}$. The indicated
errors correspond to uncertainties in the $J$ values, which were not taken
directly into account when constructing the confidence belts. Here we adopt
the central value, and note that within the quoted $J$-value
uncertainties our limits on $\lambda_{hs}$ could be weaken by up to a
factor of 1.36.

Our construction of a likelihood function for $\Phi_{\rm PP}$
works as follows. From the upper limits on $\Phi_{\rm PP}$ as a function of the
confidence level\footnote{These were kindly provided by the authors
of ref.~\cite{GeringerSameth:2011iw}.} $1-\alpha$, we determine the inverse function for the $p$-value
$\alpha=\alpha(\Phi_{\rm PP})$. Roughly speaking, this function returns the
probability (in repeated experiments) of measuring less than the observed number of events, given some true value of $\Phi_{\rm PP}$. This can be mapped onto a likelihood
function
\begin{equation}
  -2\ln\mathcal{L}_{\rm dwarfs}(\Phi_{\rm PP}) = {\rm ISF}\left[\alpha(\Phi_{\rm PP})\right]\;,
\end{equation}
where ${\rm ISF}(x)$ is the inverse survival probability function of a
$\chi^2_{k=1}$-distribution. In this way, we obtain
$-2\ln\mathcal{L}(5.0\times 10^{-30}{\rm\ cm^3s^{-1}GeV^{-2}})\simeq4.0$, as
expected for a $95\%$CL limit.
\medskip

When deriving \emph{projected} limits, we assume that \Fermi\
operates for a total of at least 10 years in the current survey mode, and that it is
able to add a further 10 new southern dwarfs to its combined search.  We
assume conservatively that the limits on $\langle\sigma v_{\rm rel}\rangle$ will scale
as $\sqrt{N}$, following the improvement in signal-to-noise ratio; our
projected \Fermi\ sensitivities are therefore based on rescaling the
current limits by a factor of $\sqrt{20/10 \times 10/3}\approx2.68$.

\subsection{CTA likelihood}
For the CTA likelihood function $\mathcal{L}_{\rm CTA}$, we reconstruct the
official CTA sensitivities for searches for dark matter annihilation towards
the Galactic Centre \cite{CTA}, with a few reasonable alternative choices for
different parameters.  Specifically, we use the ``Ring Method'', assume an NFW
\cite{NFW} DM profile, 200\,hr of observing time, and an effective area
corresponding to an extended array including both European and proposed US
contributions \cite{Jogler:2012ps}.  We include a simple background model
based on an $E^{-3}$ electron power law in the sensitivity calculation, but
neglect protons and do not consider possible systematic effects in the
background determination.  We caution that although neglecting background
systematics leads to good agreement with recent CTA projections
\cite{CTA_Snowmass}, it may result in overly optimistic sensitivities.  Full details are given in appendix \ref{ctapp}.

\subsection{Indirect detection results}

In Fig.\ \ref{fig2} we show the combined sensitivity of indirect detection to
different parts of the scalar singlet parameter space.  For current limits,
incorporating existing data from the \Fermi\ combined dwarf analysis and
WMAP7, we give only a $1\sigma$ band.  Almost no parameter space not already
excluded by relic density considerations is excluded at much higher confidence
level (CL) than this.  The region $m_h/2\le m_\S \le 70$\,GeV where $S$ makes
up all of the dark matter can be seen to be in tension with existing indirect
searches at slightly more than the $1\sigma$ level.  The same is true for a small region at $m_\S\le49$\,GeV, but this is within the area already excluded by the invisible width constraint.

Future combined limits incorporating \Planck\ polarization data, CTA
and extended \Fermi\ dwarf observations will be able to probe the
region where $S$ is all the dark matter for $m_h/2\le m_\S \le
74$\,GeV at 90\% CL.  The absence of a signal in any of these searches
will place all scalar singlet masses from $m_h/2$ to over 5\,TeV in
tension with indirect detection at more than the $1\sigma$ level, if
$S$ makes up all the DM.  As mentioned earlier however,
CTA sensitivities should be taken with something of a grain of salt. 
In Fig.\ \ref{fig3} we show the breakdown of the projected 90\% CL limit into the three different searches.  At low masses, \Fermi\ dominates the limit, whereas above $m_\S\sim m_h$, CTA takes over.  The impact of neglecting 3- and 4-body final states on the CMB limit can be seen just below $m_\S=m_\W$ and $m_\S=m_\Z$, where the CMB curve takes brief downturns before recovering once the threshold is passed.

\section{Direct detection}  
\label{dd}
We begin our discussion of the limits from direct searches with a
fresh analysis of the complementary determinations of the 
Higgs-nucleon coupling, which enters in the cross
section for singlet dark matter scattering on nuclei.  Thanks to
vigorous activity within the lattice and the theoretical
communities, this coupling seems to be better determined now than it
was just a few years ago.   For further historical details and
impacts of nuclear uncertainties on dark matter direct detection see refs.~\cite{Ellis:2008hf,
Akrami:2010dn, Bertone:2011nj}.

\subsection{Higgs-nucleon coupling}

In the past one of the largest uncertainties in the analysis of singlet DM couplings to nucleons has been the Higgs-nucleon coupling: $f_N m_N/v_0$, which depends upon the quark content of the nucleon for each quark flavour.  Here $m_N=0.946$\,GeV is the nucleon mass (we ignore the small differences between neutrons and protons here).  In general $f_N$ can be expressed in the form 
\be
f_N = \sum_q f_q = \sum_q \frac{m_q}{m_N}\langle N|\bar q q|N\rangle,
\ee
where the sum is over all quark flavours.  
The contributions from heavy quarks $q=c,b,t$ can be expressed in terms of the light ones
\be
\sum_{q=c,b,t} f_q = \frac29\left(1-\sum_{q=u,d,s}f_q\right),
\ee
by the following argument \cite{Shifman:1978zn}.  First, by equating the trace of
the stress energy tensor at low and high scales,
\be
m_N\bar N N = \sum_q m_q\bar q q - (7\alpha_s/8\pi)G_{\mu\nu}G^{\mu\nu}\;,
\ee
and taking the nucleon matrix element, one gets the relation 
\begin{equation}
m_N = m_N \sum_q f_q + \frac{21}{2}A \,,
\end{equation}
with
\begin{equation}
 A \equiv -\frac{1}{12\pi}\langle N|G_{\mu\nu}G^{\mu\nu}|N\rangle \,.
\end{equation}
Second, $\langle N|\bar qq|N\rangle$ for the heavy quarks comes from the triangle diagram that generates the $h G_{\mu\nu}G^{\mu\nu}$ coupling. Therefore the heavy-quark $f_q$ values are related to $A$ through $f_q = A/m_N$ for $q=c,b,t$.  Eliminating $A$ from these equations leads to
the claimed relation between the heavy and light quark $f_q$ values. From the above argument, the overall coupling is 
\begin{equation}
f_N = \frac29 + \frac79\sum_{q=u,d,s}f_q \,.
\end{equation}
The contributions from $u$, $d$ and $s$ are related to the light quark matrix
element $\sigma_l$ (which is related to the pion-nucleon isoscalar amplitude
$\Sigma_{\pi\rm N}$, see e.g.~ref.~\cite{Young:2013nn}): 
\begin{equation}
\sigma_l = m_l \langle N|\bar u u + \bar dd|N\rangle,
\end{equation}
where $m_l \equiv \sfrac12(m_u+m_d)$, and the non-singlet combination
\begin{equation}
\sigma_{0} = m_l \langle N|\bar u u + \bar dd - 2\bar ss|N\rangle\,,
\end{equation}
and the fairly well known isospin breaking ratio\footnote{This corrects a typo in the definition of $z$ given in ref.\ \cite{Mambrini:2011ik}.}
\begin{equation}
z = \frac{\langle N|\bar u u - \bar ss|N\rangle }{ \langle N|\bar d d - \bar ss|N\rangle} \approx 1.49 \,.
\label{isospinbreaking}
\end{equation}
In principle these relations suffice to determine all light quark $f_q$ values. Indeed, if we further define the strangeness content through the ratio
\begin{equation}
y = \frac{2 \langle N|\bar s s|N\rangle}{\langle N|\bar u u + \bar
dd|N\rangle} = 1-\frac{\sigma_0}{\sigma_l}\;,
\end{equation}
we can solve 
\begin{align}
    f_u &= {m_u\over m_u+m_d}\,{\sigma_{l}\over m_N}\,
    {2z + y(1-z)\over 1+z}\,,\nonumber\\
    f_d &= {m_d\over m_u+m_d}\,{\sigma_{l}\over m_N}\,
    {2 - y(1-z)\over 1+z}\,,\\
    f_s &= {m_\S\over m_u+m_d}\,{\sigma_{l}\over m_N}\, y\,.
    \nonumber
\end{align}
The quantities $\sigma_l$ and $\sigma_0$ have been evaluated by chiral perturbation theory (ChPT), pion-nucleon scattering and lattice simulations, with some scatter in the results. For a long time the canonical ChPT value of $\sigma_0$ was $\sigma_0 \approx 35\pm7$\,MeV~\cite{Gasser:1990ce,Gasser:1990ap,Borasoy:1996bx}, but a recent computation found $\sigma_0 \approx 58\pm9$\,MeV~\cite{Alarcon:2012kn}. Similarly, for $\sigma_l$ the older perturbation theory result was $\sigma_l \approx 45$\,MeV, whereas ref.~\cite{Alarcon:2012nr} found $\sigma_l = 59\pm7$\,MeV. The new result is in good agreement with partial wave analysis of pion-nucleon scattering ($\sigma_l = 64\pm8$\,MeV~\cite{Pavan:2001wz}), and in particular with a recent lattice evaluation ($\sigma_l = 58\pm9$\,MeV~\cite{Alvarez-Ruso:2013fza}). Depending on which of these sets one accepts, there is a wide range of possible strangeness contents of the nucleon. Fortunately, there also exist many recent, direct lattice evaluations of the strangeness matrix element:
\begin{equation}
\sigma_s = m_s \langle N|\bar s s|N\rangle \,,
\end{equation}
using 2+1 dynamical quark flavours~\cite{Bali:2011ks, Bali:2012qs,Alvarez-Ruso:2013fza,Oksuzian:2012rzb, Gong:2012nw, Freeman:2012ry,
Engelhardt:2012gd, Junnarkar:2013ac, Young:2013nn, Jung:2012rz, Gong:2013vja}. For a recent review see ref.~\cite{Young:2013nn}. Although there still is some scatter also in these results, all evaluations agree that $\sigma_s$ is quite small. Based on a subset of more constraining studies refs.~\cite{Cheng:2012qr} and \cite{Junnarkar:2013ac} reported world averages of $\sigma_s = 43\pm8$\,MeV and $\sigma_s = 40\pm10$\,MeV, respectively. However, ref.~\cite{Young:2013nn} arrived to a looser result $\sigma_s = 40\pm30$\,MeV by including also less constraining results in the analysis. (The difference between different sets may be associated with taking the correct continuum limit.)

\begin{figure*}[tb]
\includegraphics[width=1.8\columnwidth]{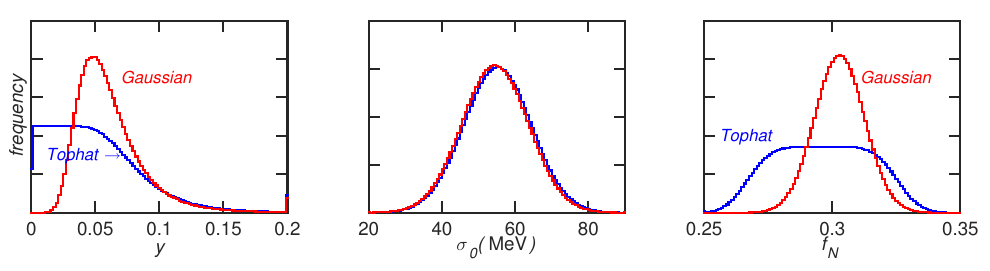}
\caption{
 Predicted distributions (in arbitrary units) of the strangeness
content $y$ of the nucleon (\textit{left}), the nucleon matrix element
$\sigma_0$ (\textit{centre}) and the Higgs-nucleon coupling factor
$f_N$ (\textit{right}).  These are drawn from a random sample
generated using experimental and theoretical constraints, as explained
in the text.}
\label{fig:FnPlots}
\end{figure*}

We have made a statistical analysis of what $f_N$ might be in light of these constraints on the nucleon matrix elements.  We choose to use the isospin breaking ratio $z$ (eq.\ \ref{isospinbreaking}) and the lattice determinations for $\sigma_l$ and $\sigma_s$ as inputs. We chose $\sigma_l$ because there is a consensus on its value when evaluated three different ways, and $\sigma_s$ because lattice simulations agree in the prediction that it is small.  To be precise, we shall use a fixed value for isospin breaking $z=1.49$ and $\sigma_l = 58\pm9$\,MeV with a Gaussian distribution. For $\sigma_s$ we explore two possibilities: either $\sigma_s = 43\pm8$\,MeV with a Gaussian distribution or $\sigma_s < 70$\,MeV with a top-hat distribution. In addition we allow the light quark masses to be Gaussian distributed with $m_q =m_{q,0} \pm \delta m_{q}$ with $\delta m_{q}\equiv \sfrac{1}{2}(\delta m_{q+}+\delta m_{q-})$ where~\cite{Cheng:2012qr}
\begin{align}
m_{u,0} &= 2.5 & \delta m_{u,+} &= 0.6 & \delta m_{u,-} &= 0.8\nonumber\\
m_{d,0} &= 5   & \delta m_{d,+} &= 0.7 & \delta m_{d,-} &= 0.9\\
m_{s,0} &= 100 & \delta m_{s,+} &= 30  & \delta m_{s,-} &= 20 \nonumber \,.
\end{align}
Here all units are in MeV. Finally, the nucleon mass is $m_N = (m_n + m_p)/2 = 938.95$\,MeV. 

With these inputs, we generate $10^7$ random realizations, from which we construct the distributions for the strangeness content $y$, the matrix element $\sigma_0$ and finally $f_N$. Results are displayed in Fig.~\ref{fig:FnPlots}. Note that $\sigma_0$ distribution is a prediction here. It is satisfying to see that it does not depend much on the strangeness input, and that the distribution ($\sigma_0 = 55 \pm 9$ MeV) agrees very well with the recent ChPT calculation ~\cite{Alarcon:2012kn}. This lends support to the self-consistency of our analysis. The strangeness content $y$  mostly reflects the input choices; the top-hat choice assumes only an upper bound for the strangeness matrix element, so $y$  is only restricted from above. This upper bound is almost the same as the upper bound in the Gaussian case, which is not consistent with $y=0$. However, what interests us is that both strangeness input choices give comparable distributions for the Higgs-nucleon coupling. In the Gaussian case we find $f_N = 0.30\pm 0.01$ at the formal 1-sigma (68.3\% CL) level. In the top-hat case we find the same mean value, but the $f_N$-distribution is broader and not Gaussian. We roughly estimate that $f_N = 0.30\pm 0.03$ in this case (see Fig.~\ref{fig:FnPlots}). Thus the error in the determination of $f_N$ is quite a lot smaller than one might believe; less than 10 per cent according to our analysis.\footnote{Note that the result quoted in the first published versions of this paper, $f_N = 0.345\pm 0.016$, was incorrect. This was due to an unfortunate, simple error in the code. All plots in this version use the corrected value.} 

\begin{figure*}[tb]
\includegraphics[width=\columnwidth]{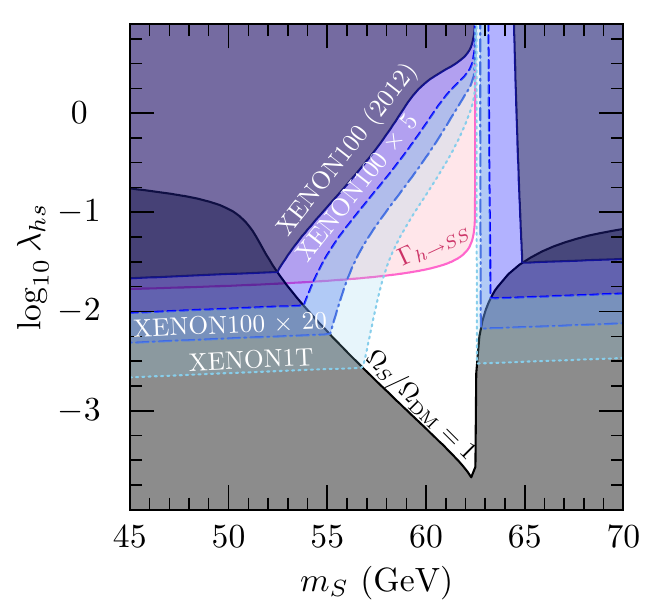}\hspace{0.06\columnwidth}
\includegraphics[width=\columnwidth]{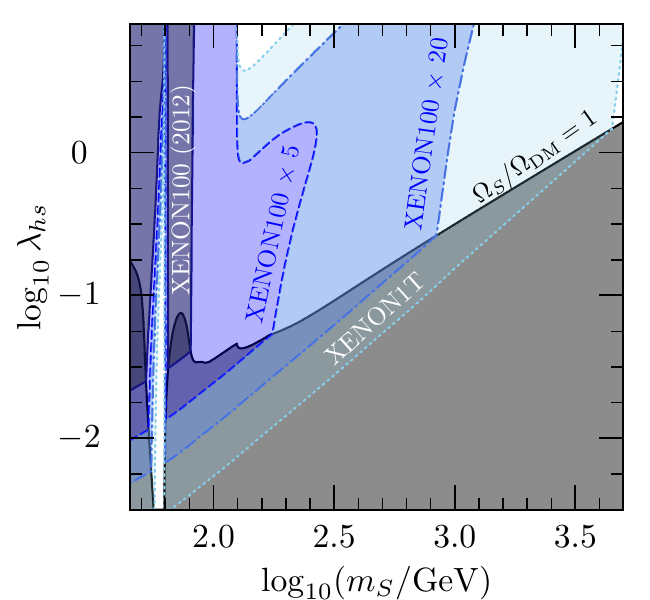}
\caption{Limits from direct detection on the parameter space of scalar singlet dark matter.  The areas excluded by present limits from 
XENON100 are delineated with solid lines and dark shading (not to be confused with the diagonal solid line and corresponding dark shading indicating the relic density bound).  Dashed, dotted and dot-dash lines indicate the areas that will be probed by future direct detection experiments, assuming 5 times the sensitivity of XENON100 (dashes, medium-dark shading), 20 times (dot-dash line, medium-light shading) and 100 times, corresponding to XENON 1-ton (dots, light shading). Note that for cases where the scalar singlet is a subdominant component of dark matter, we have rescaled the direct detection signals for its thermal relic density.  \textit{Left}: a close-up of the resonant annihilation region, with the area ruled out by the Higgs invisible width at $2\sigma$ CL indicated by the shaded region in the upper left-hand corner.  \textit{Right}: the full mass range.}
\label{ddfig1}
\end{figure*}

\begin{figure*}[tb]
\includegraphics[width=\columnwidth]{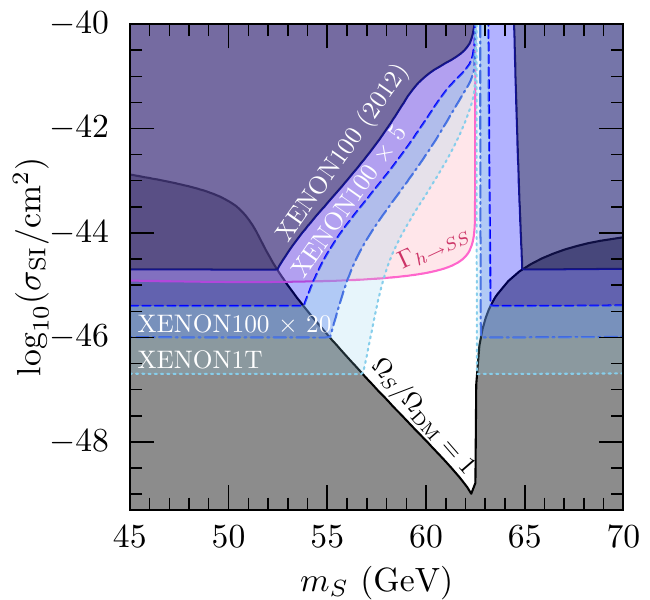}\hspace{0.06\columnwidth}
\includegraphics[width=\columnwidth]{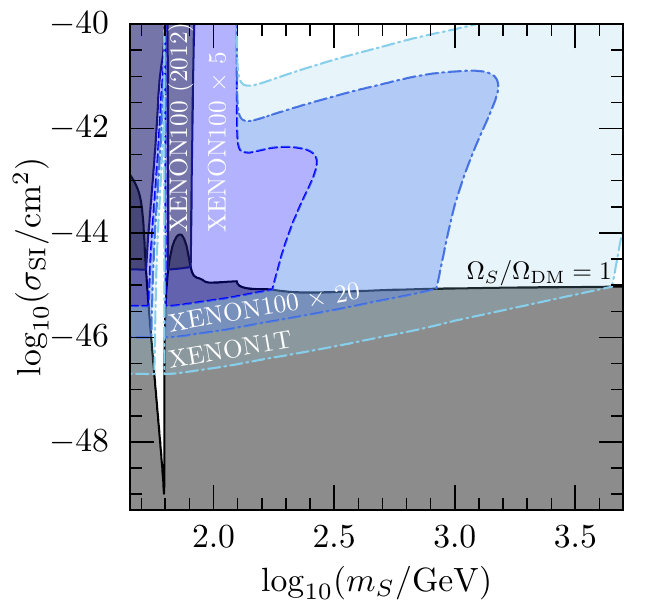}
\caption{Limits from direct detection on scalar singlet dark matter,
shown in the familiar mass-cross-section plane.  Areas excluded by
XENON100, future experiments and the relic density are as per
Fig.~\ref{ddfig1}.  The unusual shapes of the curves compared to traditional direct detection constraint plots is due to our self-consistent treatment of sub-dominant relic densities.  \textit{Left}: a close-up of the resonant annihilation region, with the area ruled out by the Higgs invisible width at $2\sigma$ CL indicated by the shaded region in the upper left-hand corner.  \textit{Right}: the full mass range.}
\label{ddfig2}
\end{figure*}

\subsection{Direct detection limits}  

The cross section for spin-independent scattering of singlet DM 
on nucleons is given by
\begin{equation}
 \sigma_{\sss\rm SI} = \frac{\lhs^2f_N^2}{4\pi}\frac{\mu^2 m^2_n}{m^4_h m^2_s},
\label{sigma_dd}
\end{equation}
where $\mu = m_n m_{\S}/(m_n+m_\S)$ is the DM-nucleon reduced mass.
The current best limit on $\sigma_{\sss\rm SI}$ comes from the XENON100
experiment \cite{Aprile:2012nq}.  In our analysis we allow for the
singlet to provide a fraction of the total dark matter, as 
indicated by the contours in Fig.\ \ref{fig:relden}.  We thus apply
the 90\% C.L.\ limits of ref.~\cite{Aprile:2012nq} (which assume a local DM density of 0.3\,GeV\,cm$^{-3}$), appropriately weighted
by the fraction of dark matter in the singlet component.  

In the
standard analysis where only a single component of DM with the full
relic density is assumed, the differential rate of detection ${\rm d}R/{\rm d}E$
is proportional to $(\rho_\odot/m_{\rm DM})\sigma_{\sss \rm SI}$, where
$\rho_\odot$ is the local DM mass density.  Thus the appropriate
rescaling of the limiting value of $\sigma_{\sss\rm SI}$ is by the fraction
$f_{\rm rel} = \Omega_\S/\Omega_{\rm DM}$ of energy density contributed by
$S$ to the total DM density.  We assume that there is no difference in
the clustering properties of the singlet component and any other component,
so that the local energy density of $S$ is $f_{\rm
rel}\,\rho_\odot$.  We therefore demand for every value of $\{\lhs,
m_\S\}$ that 
\be
\sigma_{\rm eff} \equiv f_{\rm rel}\,\sigma_{\sss\rm SI} \le \sigma_{\rm Xe} \,,
\label{sigeff}
\ee
where $\sigma_{\rm Xe}$ is the 90\% CL limit from XENON100.  Unlike with indirect signals, we do not perform this rescaling if the thermal relic density exceeds the observed value.  This is because, unlike some indirect signals, the direct detection limits depend on a mass measurement (i.e. the local density of dark matter) that is largely independent of cosmology, and therefore would not be upscaled even if the relic density were extremely large.

The resulting constraints in the $m_\S$--$\lhs$ plane are shown in
Fig.\ \ref{ddfig1}, as well as projections for how these limits
will improve in future xenon-based experiments, assuming that the
sensitivity as a function of mass scales relative to that of XENON100
simply by the exposure.  The contours showing improvements
in the current sensitivity by a factor of 5 or 20 will be relevant 
in the coming year as LUX expects to achieve such values 
\cite{Fiorucci:2013yw,Woods:2013ufa},
while XENON1T projects a factor of 100 improvement \cite{Aprile:2012zx,Beltrame:2013bba}
within two years. 
The left panel of Fig.\ \ref{ddfig1} focuses on the
resonant annihilation region $m_\S\sim m_h/2$, showing that a small
triangle of parameter space will continue to be allowed for $m_\S$ between $m_h/2$ and $\sim$57\,GeV.  Values below 53\,GeV are already robustly excluded, making it
highly unlikely that singlet dark matter can explain various hints of direct
detection that have been seen at low masses $\sim$10\,GeV
\cite{Andreas:2008xy,Andreas:2010dz}.

On the high-mass side, the right panel of Fig.\ \ref{ddfig1}
implies that most of the relevant remaining parameter space will be ruled
out in the next few years.  In particular, XENON1T will be able to
exclude masses up to 4.5\,TeV, for which the coupling must be
rather large, $\lhs>1.5$, leaving little theoretical room for 
this model if it is not discovered.

Naively, one might expect the contours of direct detection sensitivity
in the high-$m_\S$ regions to be exactly vertical in Fig.\
\ref{ddfig1} rather than being slightly inclined.   This is because
$f_{\rm eff}\sim \langle\sigma v_{\rm rel} \rangle^{-1}\sim (m_\S/\lhs)^2$ in
eq.\ (\ref{sigeff}), which is exactly inverse to
$\sigma_{\sss\rm SI}$.\footnote{There is some additional dependence upon
$\lhs$ in the annihilation cross section for $SS\to hh$, but this is
very weak at large $m_\S$.}  According to this argument, the direct
detection sensitivity would be independent of $\lhs$ and only scale
inversely with $m_\S$ due to the DM number density going as $1/m_S$.
However this is not exactly right because the DM relic density 
has an additional weak logarithmic dependence on $\langle\sigma v_{\rm rel}
\rangle$ through the freezeout temperature, leading to the relation
(see eqs.\ (\ref{Ytoday},\ref{LWeq3}), with the approximation
$A_f \cong x_f Z_f$)
\be
f_{\rm rel}\! \sim\! (x_f\, A_f)^{-1} \!\sim\! {\ln(c\,m_\S \langle\sigma v_{\rm rel}
\rangle)\over m_\S \langle\sigma v_{\rm rel} \rangle}  
\!\sim\! (m_\S \langle\sigma v_{\rm rel}\rangle)^{-1+\epsilon}\;,
\ee
for some constant $c$ and a small fractional power $\epsilon$, which we find to be
$\epsilon\cong 0.05$.  Taylor-expanding the last expression
in $\epsilon$ produces the log in the numerator.

The shape of the exclusion contours in the
$m_\S$-$\lhs$ plane of course carries over into a similar shape in 
the $m_\S$-$\sigma_{\sss\rm SI}$ plane, which is the more customary one for
direct detection constraints.  We nevertheless replot them in
this form in Fig.\ \ref{ddfig2}, to emphasize that they look very
different from the usual ones, being mostly vertical rather than horizontal.
Normally the DM relic density is assumed to take the
standard value because the annihilation cross section $\langle\sigma v_{\rm
rel} \rangle$ that sets
$\Omega_{\rm DM}$ is distinct from that for detection, $\sigma_{\sss\rm SI}$. 
Only because they are so closely related in the present model do we 
get limits that are modified by the changing relic density as one
scans the parameter space.

\begin{figure*}[tb]
\includegraphics[width=\columnwidth]{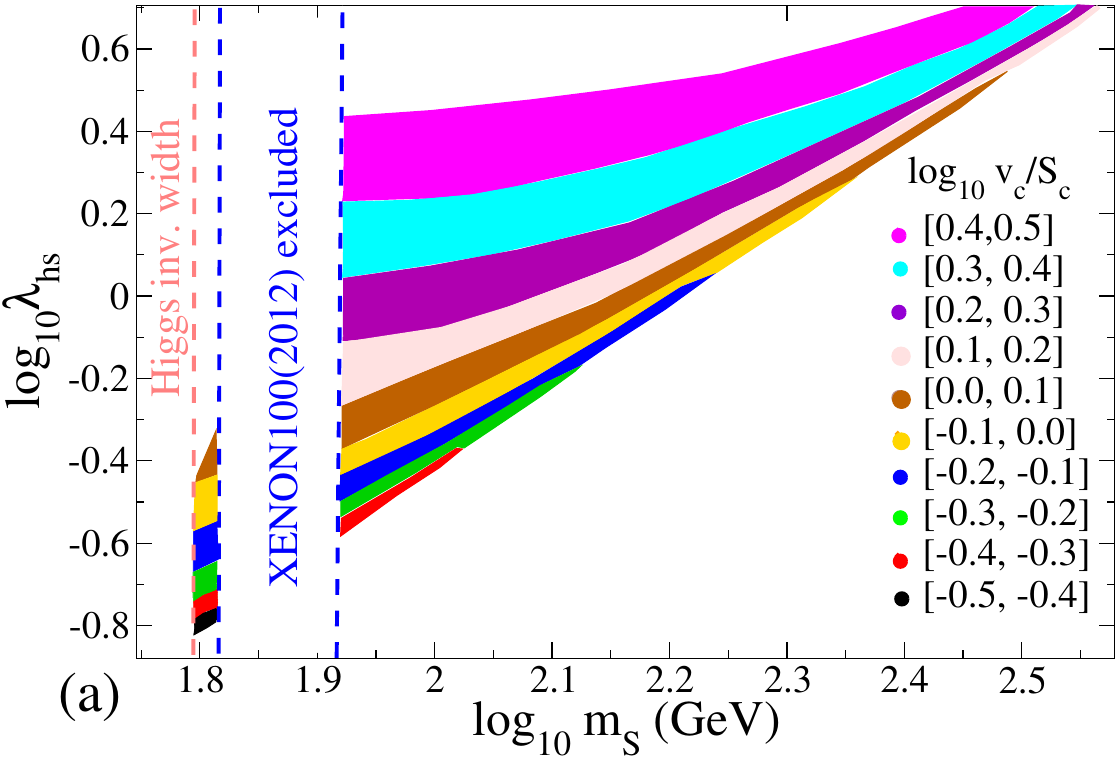}\hspace{0.06\columnwidth}
\includegraphics[width=0.98\columnwidth]{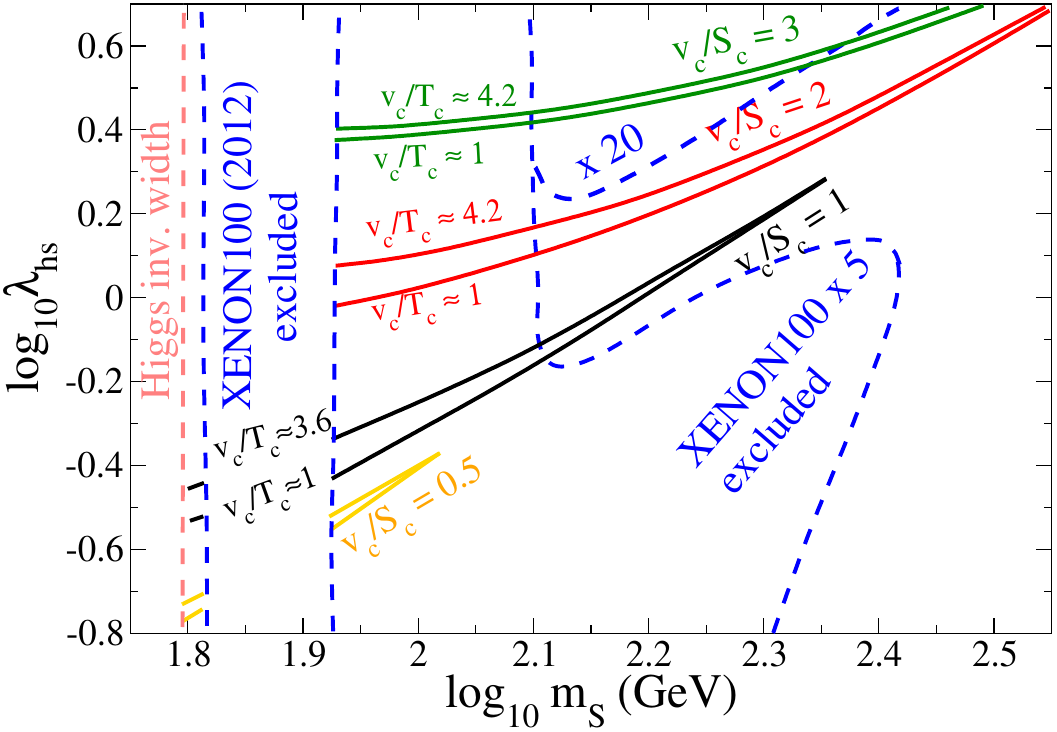}
\caption{Left: bands of models having a strong enough electroweak
phase transition for electroweak baryogenesis, scanning over the
ratio of VEVs at the critical temperature, $v_c/S_c$. Different shades
correspond to intervals of $\log_{10} v_c/S_c$ shown in the key, with
lowest values occurring lower on the plot. Right: similar
plot for fixed values of $v_c/S_c = 0.5,1,2,3$ and $v_c/T_c$ close to
1 or to its maximum value, for the given $v_c/S_c$.  The excluded region $m_\S < m_h/2$ from the
invisible Higgs width constraint is shown on the left sides of both plots.  Excluded regions
for XENON100 (2012) and for future experiments with 5 and 20 times greater
sensitivity are also shown for illustration.}
\label{vTplot}
\end{figure*}

\section{Applications}
\label{apps}

The singlet model we have considered, or modest elaborations of it,
has implications for a number of purposes other than just explaining the
dark matter, or one of its components.   These include 
strengthening the electroweak phase transition, explaining tentative
evidence for 130\,GeV and continuum gamma rays from the Galactic Centre, hints of
an extra component of dark radiation from analysis of the cosmic
microwave background, a candidate for the curvaton mechanism, and
impacting the stability of the Higgs potential near the Planck scale.
We briefly discuss these issues in the present section.

\subsection{Strong electroweak phase transition}

Recently it was pointed out that a strong electroweak phase 
transition (EWPT), with $v_c/T_c\ge 1$ at the critical temperature,
can be obtained in the scalar singlet dark matter model if
$\lhs\gtrsim 0.1$ \cite{Cline:2012hg}, thus requiring the singlet to 
comprise a sub-dominant component of the total dark matter density.
The criterion $v_c/T_c>1$ is needed for a successful model of electroweak 
baryogenesis (also considered in ref.~\cite{Cline:2012hg}).  The effect of
the singlet on the EWPT depends upon an additional operator
$\lambda_\S S^4$ which was not relevant for the preceding analysis.
By scanning over $\lambda_\S$, ref.\ \cite{Cline:2012hg} produced 
many random realizations of models giving a strong enough EWPT.
Here we have repeated this procedure in order to display the range of
viable models in the space of $\{m_\S, \lhs\}$ for comparison with
figs.\ \ref{fig:relden}-\ref{ddfig2}.  

In these models, the $Z_2$
symmetry $S\to -S$ is temporarily broken by a VEV $S_c$ at the
critical temperature.  It is convenient to parametrize the $S^4$
coupling as $\lambda_\S = (\lambda_h/4)(v_c/S_c)^4$ where $\lambda_h
=0.13$ is the Higgs quartic coupling.  We consider $(v_c/S_c)^4$ in the
range $0.1-10$, corresponding to $\lambda_S \in [3\times10^{-4},3]$.
The results are shown in Fig.\ \ref{vTplot}.  
In the left panel,
shaded bands of models correspond to intervals of $v_c/S_c$ as shown
in the key of the figure; larger $v_c/S_c$ corresponds to larger
$\lhs$ at a given mass $m_\S$.  
There is an island of small $\lhs$
near $m_\S\sim m_h/2$ where $SS$ annihilations are resonantly
enhanced.  These correspond to $v_c/S_c < 1$.  In the right panel, we
take several discrete values of $v_c/S_c$ to better illustrate the
dependence of $v_c/T_c$ on the parameters $m_\S,\lhs$.  For a given
value of $v_c/S_c$, there is always a maximum mass $m_\S$ beyond which
there is no longer a strong phase transition.  For large $v_c/S_c$,
this occurs at strong couplings $\lhs > 5$ that we do not consider.

Contours showing the current direct detection limit \cite{Aprile:2012nq} 
and projected ones for experiments with 5 and 20$\times$ greater
sensitivity are also shown in the right panel of Fig.\ \ref{vTplot}.
A large region 
of the remaining parameter space will be excluded by 
the LUX experiment \cite{Fiorucci:2013yw,Woods:2013ufa}, which plans
to achieve a factor of better than
10$\times$ improvement relative to ref.\ \cite{Aprile:2012nq} by the end of 2013.  Within two years, XENON1T
expects to reach 100$\times$ the sensitivity of the XENON100 (2012)
\cite{Aprile:2012zx,Beltrame:2013bba}.

\begin{figure}[tb]
\includegraphics[width=\columnwidth, trim = 0 0 8 0, clip=true]{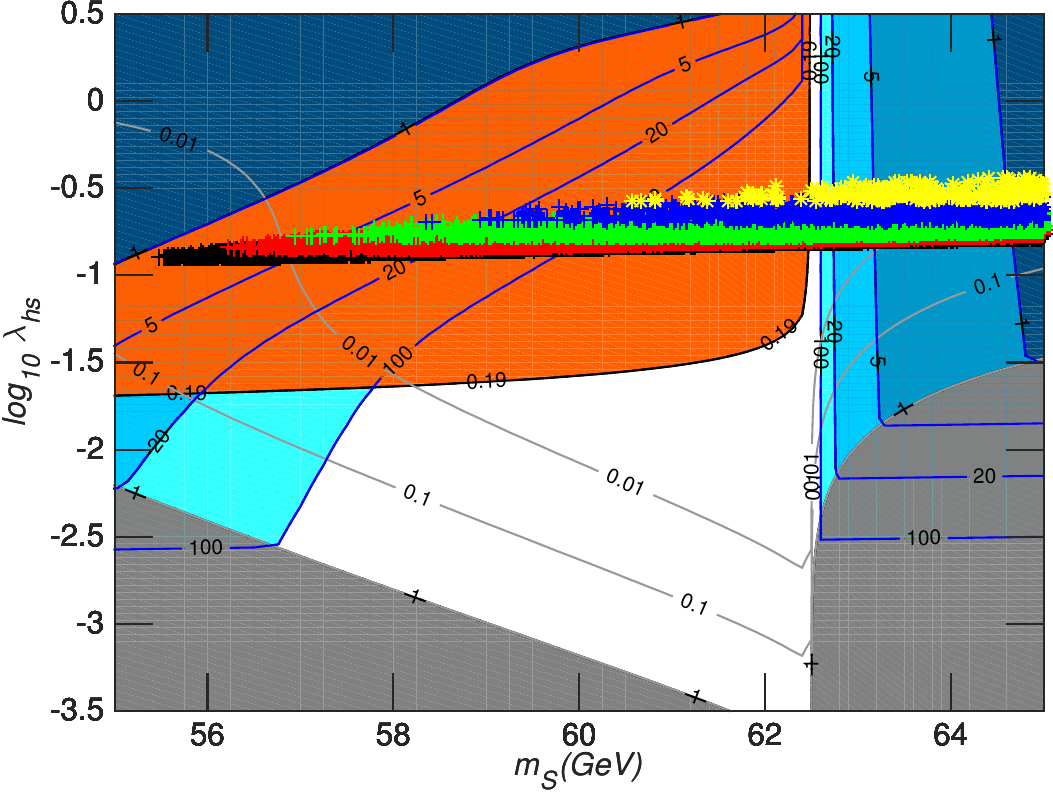}
\caption{Scatter plot of models with strong EWPT, focusing on the
low-mass region near $m_h/2$.  Shading of points follows Fig.\ \ref{vTplot}.  Limits from XENON100, and from future experiments with 5 and $20\times$ greater sensitivity, are shown as vertical lines to the right of the plot and diagonal lines to the left, with the ruled out areas marked by graded (blue) shading.  The area ruled out by the Higgs invisible width at $2\sigma$ CL lies above and to the left of the line labelled ``$\Gamma_{h\to SS}$''.  The area ruled out by the relic density constraint is shown as usual as a dark shaded region at the bottom of the plot, with additional labelled contours indicating lines of constant sub-dominant relic density.}
\label{vTplot2}
\end{figure}

The island of models near $m_\S\sim m_h/2$ is squeezed on the left
by the requirement $m_\S > m_h/2$ due to the constraint on the
invisible width of the Higgs, and on the right by the direct detection
bound.  This region will become increasingly narrow as the XENON
bounds improve, as shown close-up in Fig.\ \ref{vTplot2}. 

\subsection{130\,GeV gamma-ray line}

There has been significant interest in tentative evidence for a 130
GeV gamma-ray line from the Galactic Centre found in Fermi-LAT data
\cite{Bringmann:2012vr, Weniger:2012tx, Tempel:2012ey, Su:2012ft,
Hektor:2012ev, Finkbeiner:2012ez, Whiteson:2012hr, Fermi-LAT:2013uma}, which might be
interpreted as coming from annihilation of dark matter.  In ref.\
\cite{Cline:2012nw} it was suggested that the scalar singlet dark
matter model could provide an explanation, if one added an
additional interaction $\lambda_{S\sigma}S^2|\sigma|^2$ 
with a charge-two singlet $\sigma$, transforming in the fundamental
representation of a new SU(N) gauge interaction.  Then $SS$ can
annihilate into $\gamma\gamma$ through a virtual loop of $\sigma$,
producing gamma rays of the observed energy if $m_\S = 130$\,GeV.

To get a large enough cross section into photons, $S$ should be the
dominant dark matter particle, hence $\lhs$ should be close to $0.05$.
From the right panel of Fig.\ \ref{ddfig1} and the previous discussion, it is clear that these values will be probed 
in the coming year by LUX.  This conclusion could be evaded if
glueballs of the new SU(N) are lighter than 130\,GeV however; in that
case $\lhs$ could be much less than $0.05$ to evade the direct
detection limit, while the $S$ relic density could be achieved by
annihilation of $SS$ into glueballs, via the $\sigma$ loop.

\subsection{Continuum gamma rays from the Galactic Centre}

An excess of continuum gamma rays has also been claimed in
Fermi-LAT data towards the Galactic Centre \cite{Hooper:2010mq,
Hooper:2011ti, Abazajian:2012pn, Hooper:2013rwa}.  This has been interpreted as
consistent with annihilation of dark matter with a mass of 
$30-50$\,GeV and a cross-section of $\langle \sigma v_{\rm rel} \rangle_0
\sim 6-8 \times 10^{-27}$\,cm$^3$\,s$^{-1}$ into $b$ quarks
\cite{Hooper:2011ti,Hooper:2013rwa}.  Considering that the Fermi-LAT dwarf limit on
annihilation into $b\bar b$ is $\langle \sigma v_{\rm rel} \rangle_0
\le 4 \times 10^{-26}$\,cm$^3$\,s$^{-1}$ at a mass of 50\,GeV
\cite{GeringerSameth:2011iw,Ackermann:2011wa}, and remembering that
$\sigma v_{\rm rel}$ scales roughly as $\lhs^2$ for fixed $m_\S$, we
see that all models that could approximately fit this signal (i.e. with appropriate cross-sections and masses below $\sim$60\,GeV) lie less than an order of magnitude above the indirect limit shown in Fig.\
\ref{fig2}.  At low masses, all these models are therefore excluded by
the Higgs invisible width, and above 53\,GeV, their thermal relic
densities all grossly exceed the observed cosmological abundance of
dark matter.  Scalar singlet dark matter therefore cannot be
responsible for the observed continuum gamma rays at the Galactic
Centre, unless the theory is supplemented by some additional physics
that would suppress the thermal relic density.

\subsection{Complex singlet dark matter}
Another natural generalization of scalar singlet dark matter is
the case where $S$ is a complex scalar.  With no additional
interactions, this would be equivalent to two real singlets, and the
potential is most naturally written in the form
\be
    V =  \mu_S^2 |S|^2 + \lhs |S|^2|H|^2\;,
\label{spot2}
\ee
with $S = (S_1 + i S_2)/\sqrt{2}$ giving the relation to the
canonically normalized real singlets $S_{1,2}$.  The relic density
$n$ would thus be doubled relative to the real singlet model with the
same values of $m_\S$ and $\lhs$, and since 
$n$ scales as $1/\langle\sigma v_{\rm rel}\rangle \sim \lhs^{-2}$, our relic
density contours would thus move upward by 
$\delta\log_{10}\lhs \cong 0.15$.  The direct detection signal scales
roughly as $N/m_\S$ for $N$ components of degenerate dark matter, so
the contours for direct detection would move to the right by $\delta
\log_{10}m_\S\cong 0.3$. 

It was recently suggested that hints from the CMB of an extra
component of dark radiation could be explained in the context of 
fermionic singlet dark matter if the U(1) symmetry $\psi\to
e^{i\alpha}\psi$ for dark matter number conservation is spontaneously
broken near the weak scale.  This  leads to Nambu-Goldstone bosons
comprising the dark radiation, and a small mass splitting between the
two dark matter components \cite{Weinberg:2013kea}.  Scalar singlet
dark matter as we consider here offers an alternative implementation 
of this idea; by adding an extra scalar $X$ that carries dark matter
charge $1$ or $2$ and whose potential gives it a VEV, we can achieve a
similar result.  We leave the details for future investigation.

\subsection{Curvaton model}
The same model as we are studying as a dark matter candidate has
recently  
been considered as a curvaton candidate in ref.\ 
\cite{Enqvist:2012tc}.  The curvaton is a massive field whose fluctuations
during inflation later come to dominate the universe, before they
decay and produce the primordial density fluctuations.  This is an
interesting alternative to inflaton fluctuations in the case where the
latter are sub-dominant.  In the present model, $S$ cannot decay, but
its annihilations through resonant preheating can convert its
fluctuations into Higgs particles which then decay into other standard
model particles.

The region of interest in the parameter space $\{m_\S,\lhs\}$
considered by ref.\ \cite{Enqvist:2012tc} is $m_\S\in [10^2,10^{11}]$
GeV, $\lhs\in 2\times[10^{-2},10^{-30}]$, which according to our
analysis should be entirely ruled out.  However we have assumed that
the dark matter thermalizes at high temperatures and freezes out in
the standard way, whereas the curvaton decay process is a non-thermal
one, which can only be reliably calculated until the not-too-late 
stages
of preheating.  If the universe thermalizes in this scenario to a
maximum temperature below the standard freeze-out value for the dark
matter, then it is possible that $S$ could be the curvaton and evade
our constraints, while possibly even attaining the right relic density
through this non-thermal mechanism.  However it would be numerically
very challenging to test the scenario given the current
limitations of lattice codes for preheating.

\subsection{Higgs potential stability}
A curious feature of the recently determined value of the Higgs boson
mass is that it is slightly below what would be needed to maintain
positivity of the quartic Higgs coupling $\lambda_h$ under
renormalization group running up to the Planck scale 
assuming only
the standard model \cite{Degrassi:2012ry}.  The top quark gives a
large negative contribution to the running of $\lambda_h$, which is
not quite offset by the positive contribution from $\lambda_h$ itself.
However the coupling $\lhs$ gives an additional positive contribution
which has the potential to bring about stability of $\lambda_h$.
This effect has been previously studied in refs.~\cite{Lerner:2009xg,Gonderinger:2009jp,Kadastik:2011aa,Barger:2008jx}.

Although higher order corrections are needed to make an accurate
prediction, one can reasonably approximate the size of the effect using
 the one-loop contributions to
the beta function $\beta_{\lambda_h}$, in order to make a rough
estimate of the magnitude of $\lhs$ needed in order to have an impact 
on the vacuum stability question.  It was shown in ref.~\cite{Degrassi:2012ry}
that a shift in the top quark mass $\delta m_t = -2$\,GeV would be
sufficient to yield positivity of $\lambda_h$ up to the Planck scale
for $m_h = 125$\,GeV.  This corresponds to a shift in $\beta_{\lambda_h}$
of \cite{Sirlin:1985ux}
\be
    \delta\beta_{\lambda_h} = -24 {\delta m_t\over m_t}
    {y_t^3\over 16\pi^2} \cong {0.28\over  16\pi^2},
\ee
where $y_t$ is the top quark Yukawa coupling.
On the other hand, the scalar singlet contributes an amount
\be
    \delta\beta_{\lambda_h} = {\sfrac12\lhs^2\over 16\pi^2}.
\ee
According to this estimate, values near $\lhs\sim 0.75$ could be
sufficient to achieve stability of the Higgs potential, which would
correspond to DM masses $m_\S\sim 3$ TeV.  

The previous argument ignores the
effect of the $\lambda_\S S^4$ coupling on the running of 
$\lambda_h$, which was shown in ref.\ \cite{Gonderinger:2009jp} to reduce the
effectiveness of $\lhs$ for improving vacuum stability.  Inspection
of their results (see Fig.\ 1 of ref.~\cite{Gonderinger:2009jp}) confirms the above 
estimate for the needed size of $\lhs\sim 0.75$.

\section{Conclusions}
\label{summary}

The model of scalar singlet dark matter $S$ was proposed at least 28
years ago.  We have reconsidered the prospects for its discovery by
direct or indirect signals and found that the next two years are
likely to be crucial.  In particular the XENON1T experiment should
discover or rule out the scalar singlet for most reasonable values of
its mass and coupling $\lhs$ to the Higgs, leaving only values
$\lhs>1.5$ that start to be non-perturbative. We find that in a small 
range of masses $m_S \sim 57-62.5$ GeV and couplings 
$-2 \gtrsim \log_{10} \lambda_{sh} \gtrsim -3.5$ 
the singlet scalar DM cannot be ruled out by any of the forthcoming observations. 
However in this region our momentum-independent relic density 
calculation, which solves only for the abundance rather than the DM
distribution function, should be verified by use of a full 
momentum-dependent Boltzmann code.  We argued that the theoretical
uncertainty in the Higgs-nucleon coupling, which has long affected
predictions, is now significantly smaller than it was until only 
rather recently.

If the model is excluded by direct searches then constraints from
indirect detection will no longer be competitive, but the situation
will be more interesting if there is a direct detection.  In that
case, complementary information will be required to see whether the
singlet model is preferred over other possible models.  We have shown
(Fig.\ \ref{fig3}) that there is a region of parameter space where $S$
provides a not-too-small fraction of the total dark matter while still
giving an observable signal in gamma rays that might be detected by
the \v{C}erenkov Telescope Array.  Interestingly, this includes a theoretically
motivated region where the singlet's effect on the running of the
Higgs self-coupling $\lambda_h$ could push it 
back to a positive value at the Planck scale.

Unfortunately, for most values of the mass $m_\S$, there is typically
a rather large range of values of its coupling $\lhs$ to the Higgs 
for which direct detection would be possible, but not indirect
detection.  These include the regions where $S$ could help to induce a
strong electroweak phase transition.  The prospects for indirect
detection would be dramatically improved if $S$ couples to some new
charged particles, which has been suggested as a scenario for
explaining hints of 130\,GeV dark matter annihilating into gamma rays at the Galactic
Centre.  This intriguing possibility too will be settled in the near
future, both by improvements in direct detection sensitivity, and
imminent observations by the HESS-II experiment 
\cite{Bergstrom:2012vd}.

\acknowledgments{We thank Alex Geringer-Sameth for kindly providing
his results from ref.\ \cite{GeringerSameth:2011iw} at arbitrary confidence
levels, Jan Conrad, Jenny Siegal-Gaskins and Martin White for helpful
discussions on CTA, and Ankit Beniwal for picking up an earlier numerical error in the calculation of $f_N$.
J.C.\ is supported by the Natural Science and
Engineering Research Council of Canada (NSERC), and thanks the University of Jyv\"askyl\"a Physics
Department for its hospitality while this work was being completed.  P.S.\ is supported by
the Banting program, administered by NSERC.}

\appendix
\section{$s$-dependent cross-sections}
\label{exact x-sections}
As explained in the main text, we cannot use eq.~(\ref{sv}) with the tabulated values of ref.~\cite{Dittmaier:2011ti} for Higgs boson widths $\sqrt{s}\gtrsim 300$\,GeV. Instead, we have to use the perturbative cross sections into kinematically open channels, which are dominated by the gauge bosons and the top quark. The cross section into gauge bosons is:
\begin{equation}
v_{\rm rel}\sigma_{\rm VV} = \frac{\lambda_{hs}^2s}{8\pi}\delta_Vv_V|D_h(s)|^2
(1 - 4x + 12x^2)\,,
\label{GGx-section}
\end{equation}
where $x\equiv M_V^2/s$, $v_{\rm V} = \sqrt{1-4x}$ and $\delta_W = 1$, $\delta_Z = \sfrac{1}{2}$
and $|D_h(s)|^2$ is defined in eq.~(\ref{eq:Dh}).
Annihilation into fermion final states is given by
\begin{equation}
v_{\rm rel}\sigma_{\rm f\bar f} = \frac{\lambda_{hs}^2m_{\rm f}^2}{4\pi} X_{\rm f} v_{\rm f}^3|D_h(s)|^2 \,,
\label{ffx-section}
\end{equation}
where $v_{\rm f} = \sqrt{1-4m_{\rm f}^2/s}$ and $X_{\rm f} = 1$ for leptons, while for quarks it incorporates a colour factor of 3 and an important one-loop QCD correction~\cite{Drees:1990dq}:
\begin{equation}
X_q = 3 \left[1 + \left( \frac{3}{2}\log\frac{m_q^2}{s} + \frac{9}{4}\right) \frac{4\alpha_s}{3\pi} \right] \,,
\end{equation}
where $\alpha_s$ is the strong coupling for which we take the value $\alpha_s= 0.12$. Using QCD-corrected annihilation rates for light quarks is an excellent approximation below the lower limit $\sqrt{s} = 90$ GeV to which ref.~\cite{Dittmaier:2011ti} gives tabulated results. Neglecting QCD-corrections there would lead to an error of order ${\cal O}(1)$. Of course this region turns out to be ruled out. In the large mass region the QCD correction on the top-quark final state is quite small.

Finally the annihilation cross section to the Higgs boson pairs is given by
\begin{eqnarray}
v_{\rm rel}\sigma_{hh} 
&=& \frac{\lambda_{hs}^2}{16\pi s^2 v_\S}
    \left[ \; (a_R^2 + a_I^2) s v_\S v_h    \phantom{\frac{1}{2}}    
    \right. 
\nonumber \\
&+&  \left. 4\lambda_{sh} v_0^2 \left( a_R - \frac{\lambda_{sh}v_0^2}{s-2m_h^2} \right)
          \log \left| \frac{m_\S^2 - t_+}{m_\S^2 - t_-} \right|  \right. 
          \nonumber \\
& & \left. \hskip-0.4cm
        + \frac{2 \lambda_{sh}^2v_0^4 s v_\S v_h}{(m_\S^2 - t_-)(m_\S^2 - t_+)} \; 
    \right] \,,
\label{svhh}
\end{eqnarray}
where $v_i = \sqrt{1 - 4m_i^2/s}$,  $t_\pm =  m_\S^2 + m_h^2 - \sfrac{1}{2}s (1  \mp  v_\S v_h)$, and 
\begin{eqnarray}
a_R  &\equiv& 1 + 3 m_h^2(s-m_h^2)|D_h(s)|^2 \nonumber \\
a_I  &\equiv& 3m_h^2\sqrt{s}\,\Gamma_h(m_h)|D_h(s)|^2.
\end{eqnarray}
In the zero-velocity limit $\sqrt{s}=2m_\S$ this cross section immediately reduces to the expression given in eq.~(4.1) of ref.~\cite{Cline:2012hg}.

%
\section{Solution of the Boltzmann equation}
\label{appA}
%

{The Lee-Weinberg equation for the number density can be written as
\begin{equation}
\frac{{\rm d}Y}{{\rm d}x} = Z(x) \left[Y_{\rm eq}^2(x) - Y^2(x) \right]\,,
\label{LWeq}
\end{equation}
where $Y \equiv n/s$ is the ratio of the WIMP number density $n$ to entropy $s$, $x\equiv m/T$ and 
\begin{equation}
Z(x) \equiv \sqrt{\frac{\pi}{45}}\frac{m_S M_{\rm Pl}}{x^2} [ \sqrt{g_*}\langle v_{\rm rel} \sigma \rangle](x) \,,
\end{equation}
where the average cross section $\langle v_{\rm rel} \sigma\rangle$ is given in
eq.~(\ref{thermsv}) and
\begin{equation}
\sqrt{g_*} \equiv \frac{h_{\rm eff}}{\sqrt{g_{\rm eff}}}
\left( 1 + \frac{T}{3h_{\rm eff}} \frac{{\rm d}h_{\rm eff}}{{\rm d}T} \right)\,,
\end{equation}
where $h_{\rm eff}$ and $g_{\rm eff}$ are the effective entropy and energy
degrees of freedom, which we compute assuming standard model particle content.
Finally,
\begin{equation}
Y_{\rm eq}(x) = \frac{45}{4\pi^4} \frac{x^2}{h_{\rm eff}(T)}K_2(x)
\end{equation}
in the Maxwell-Boltzmann approximation. We solve eq.~(\ref{LWeq}) both
numerically and in a semi-analytic freeze-out approximation, which
differs slightly from the one usually presented in the
literature~\cite{Gondolo:1990dk,Kolb:1990vq}. For a similar treatment see
however ref.~\cite{Steigman:2012nb}. We begin by defining $Y \equiv (1+\delta) Y_{\rm eq}$ and rewriting the Lee-Weinberg equation as an equation for $\delta$:
\begin{equation}
\frac{{\rm d}\delta}{{\rm d}x} + (1+\delta) \frac{{\rm d}\log Y_{\rm eq}}{{\rm d}x} = -Z(x)Y_{\rm eq}(x) \delta(\delta + 2) \,.
\label{LWeq2}
\end{equation}
The freeze-out approximation is based on the observation that $\delta$ starts
to grow slowly, such that ${\rm d}\delta/{\rm d}x \ll \delta$ until $\delta
\sim {\cal O}(1)$.\footnote{Note that due to the leading exponential behaviour at large $x$ ${\rm d} \log Y_{\rm eq}/{\rm d}x \approx -1$.} When this holds, one can neglect the $\delta$-derivative and reduce (\ref{LWeq2}) into an algebraic equation for $\delta=\delta(x)$. We turn this argument around by assuming that the approximation holds until some freeze-out value $\delta_{\rm f}$ and solve the corresponding freeze-out $x_{\rm f}=x(\delta_{\rm f})$ from the ensuing condition:
\begin{equation}
x_{\rm f} = \log \left( \frac{ \delta_{\rm f}(2+\delta_{\rm f} ) }{ 1+\delta_{\rm f} } 
\frac{ Z \hat Y_{\rm eq}^2 }{\hat Y_{\rm eq} - \frac{{\rm d}\hat Y_{\rm eq}}{{\rm d}x}} \right)_{x_{\rm f}} \,, 
\label{xfout}
\end{equation}
where  $\hat Y_{\rm eq} \equiv  e^{x}Y_{\rm eq}$. Equation (\ref{xfout}) is simple to solve by iteration. At $x=x_f$ one then has $Y_{\rm f} = (1+\delta_{\rm f})Y_{\rm eq}(x_{\rm f})$. For $x>x_{\rm f}$ one may safely neglect the $Y^2_{\rm eq}$-term (back reaction), which allows us to integrate the equation exactly to the final result:
\begin{equation}
Y_{\rm today} = \frac{Y_{\rm f}}{1 + Y_{\rm f} A_{\rm f}} \,, 
\label{Ytoday}
\end{equation}
where
\begin{equation}
A_{\rm f} = \int_{x_{\rm f}}^\infty {\rm d}x Z(x) \,.
\label{LWeq3}
\end{equation}
The $A_{\rm f}$-integral is easy to do numerically. We show the
comparison of the numerical and the freeze-out solution of the
Lee-Weinberg equation (\ref{LWeq}) in Fig.~\ref{fig:freezeout} for
$\lhs=1$ and $\delta_{\rm f}=1$. 
Overall, the freeze-out approximation (\ref{Ytoday})  is found to be
accurate to $0.3\%-0.7\%$ over most of the parameter space in our
model, the exception being close to the Higgs resonance where the
error can reach $1.7$\%. The dependence of the freeze-out solution on $\delta_{\rm f}$
is at sub-per cent level for $\delta_{\rm f}= 0.5-1.5$. Let us point
out that if the the quantity $\sqrt{g_*}\langle v_{\rm rel} \sigma
\rangle $ is weakly dependent on $x$, one can approximate $A_{\rm f}
\approx x_{\rm f} Z_{\rm f}$. This approximation is typically accurate
to a few per cent at large masses and away from resonances, but it
becomes much less accurate near resonances or places where
$\sqrt{g_*}\langle v_{\rm rel} \sigma \rangle$ has abrupt features as
a function of $x$.

\begin{figure}[tb]
\includegraphics[width=0.9\columnwidth]{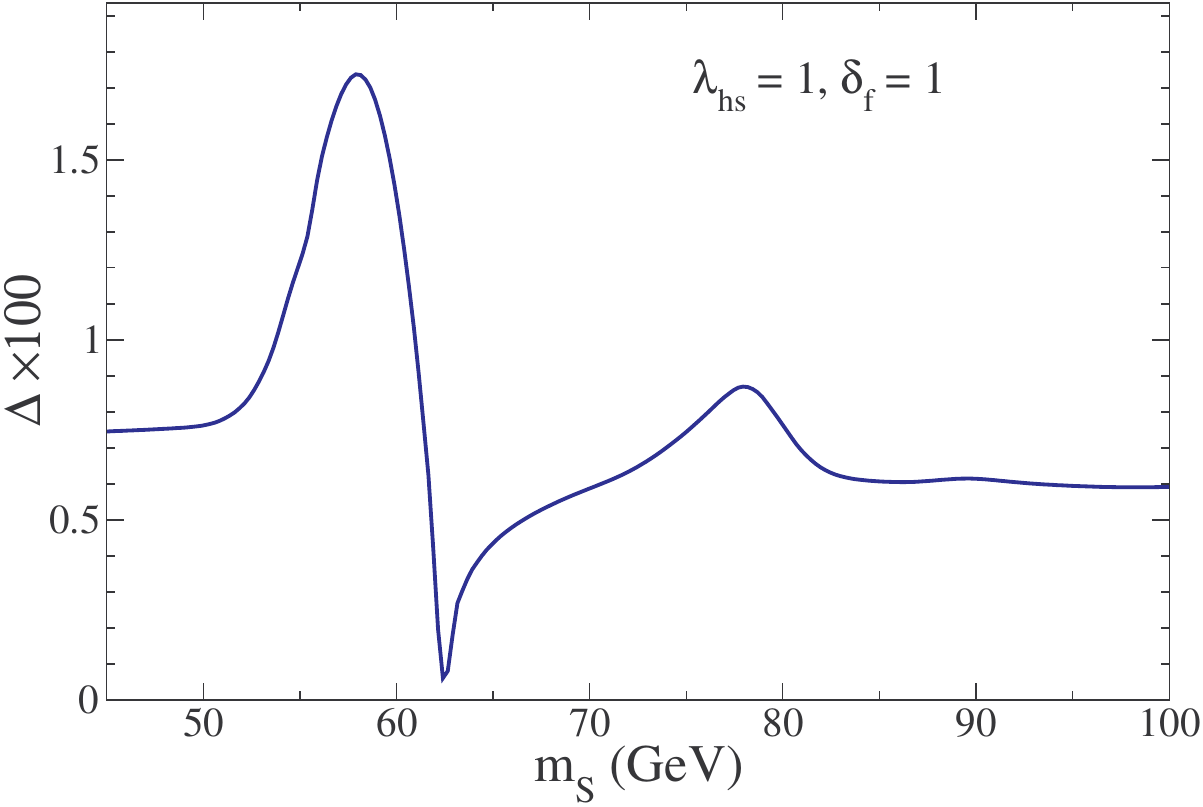}
\caption{Shown is the relative difference $\Delta \equiv (Y^{\rm full}_{\rm today} - Y_{\rm today})/Y_{\rm today}$ where $Y_{\rm today}$ if obtained from 
eq.~(\ref{Ytoday}) and $Y^{\rm full}_{\rm today}$ from a direct numerical integration of the Lee-Weinberg equation.}
\label{fig:freezeout}
\end{figure}

\section{CMB $\boldsymbol{f_{\rm eff}}$ at WIMP masses above 1\,TeV}
\label{cmbappendix}
As a supplement to the results of ref.~\cite{Cline:2013fm}, in Table\ \ref{tab1} we give values of $f_{\rm eff}$ for WMAP7 and Planck at WIMP masses $m_\chi$ of 3\,TeV and 10\,TeV.

\begin{table}[ht]
\begin{tabular}{|c||c|c||c|c|}
 \hline
   $m_\chi\to$ &   3\,TeV &   10\,TeV &  3\,TeV &  10\,TeV \\
\hline
channel & \multicolumn{2}{|c||}{WMAP7 $f_{\rm eff}$} &
\multicolumn{2}{|c|}{Planck $f_{\rm eff}$} \\
 \hline
 $e$        &  0.55 &  0.55 & 0.59  & 0.58 \\
 \hline                                    
 $\mu$      &  0.21 &  0.21 & 0.22  & 0.22 \\
 \hline                                    
 $\tau$     &  0.18 &  0.18 & 0.19  & 0.19 \\
 \hline                                    
 $V\to e$   &  0.56 &  0.56 & 0.60  & 0.60 \\
 \hline                                    
 $V\to\mu$  &  0.20 &  0.20 & 0.21  & 0.21 \\
 \hline                                    
 $V\to\tau$ &  0.18 &  0.18 & 0.19  & 0.19 \\
 \hline                                    
 $q(u,d,s)$ &  0.26 &  0.26 & 0.28  & 0.28 \\
 \hline                                    
 $c$        &  0.27 &  0.26 & 0.28  & 0.28 \\
 \hline                                    
 $b$        &  0.27 &  0.26 & 0.28  & 0.28 \\
 \hline                                    
 $t$        &  0.25 &  0.25 & 0.27  & 0.26 \\
 \hline                                    
 $\gamma$   &  0.54 &  0.52 & 0.57  & 0.56 \\
 \hline                                    
 $g$        &  0.27 &  0.26 & 0.28  & 0.28 \\
 \hline                                    
 $W$        &  0.24 &  0.24 & 0.25  & 0.26 \\
 \hline                                    
 $Z$        &  0.22 &  0.22 & 0.23  & 0.23 \\
 \hline                                    
 $h$        &  0.25 &  0.24 & 0.27  & 0.26 \\
 \hline
\end{tabular}
\caption{$f_{\rm eff}$ values for WIMP masses $m_\chi$ above 1\, TeV, in different primary
annihilation channels, for computing WMAP7 (left) and projected Planck 
(right) constraints.  As an example, ``$\mu$'' denotes $\chi\chi\to 
\mu\bar \mu$, whereas ``$V\to \mu$'' denotes $\chi\chi\to VV$, followed by 
$V\to \mu\bar \mu$.  See ref.~\protect\cite{Cline:2013fm} for further details.}
\label{tab1}
\end{table}

\section{CTA likelihood details}
\label{ctapp}

We use the Ring Method as outlined in ref.\ \cite{CTA}, as optimized for CTA
candidate Array B.  The Ring Method is an advanced version of the standard
ON-OFF analysis, where the telescope is pointed slightly away from the
Galactic Centre (GC), and the ON region (called the ``signal region'' in the
Ring Method although it may contain both signal and background) and OFF region
(called the ``background region'' although it may also contain both signal and
background) are defined as different portions of a ring centred on the centre
of the field of view.  A band covering the Galactic plane is excluded from
both the signal and background regions.  We calculate the signal and
background-region line-of-sight integrated $J$ factors for DM annihilation
towards the GC assuming the NFW profile of ref.~\cite{NFW,Battaglia:2005rj}
(namely, a local density of $0.29\rm GeV/cm^3$ and a scale radius of
$r_s=17\rm\,kpc$) and a moderate substructure boost factor of around 3, obtaining $J_{\rm ON} = 6.6\times10^{21}$\,GeV$^2$\,cm$^{-5}$ and $J_{\rm OFF}=7.7\times10^{21}$\,GeV$^2$\,cm$^{-5}$.  Even with this mild boost, our signal(ON)-region $J$ factor is still approximately a factor of 6 smaller than given in ref.\ \cite{CTA}, most likely because the density profile used in ref.\ \cite{CTA} was based on the Aquarius $N$-body simulation \cite{Springel:2008cc} rather than stellar kinematic fits.

In the absence of any publicly-available effective area corresponding to Array B, we use the energy-dependent effective area $A_{\rm eff}(E)$ given for an extended array in ref.\ \cite{Jogler:2012ps}.  This effective area corresponds to a European baseline array of 25 medium-sized Davis-Cotton telescopes, plus an additional (less likely) proposed US contribution of 36 medium-sized Davis-Cotton telescopes.  The expected number of events in the observable energy window (approximately 30\,GeV--8\,TeV for this array) is then
\begin{align}
\theta_k &= \theta_{k,{\rm BG}} + \theta_{k,{\rm DM}} \nonumber\\
         &= \theta_{k,{\rm BG}} + t_{\rm obs} J_k \frac{\langle\sigma v_{\rm
rel}\rangle}{8\pi m_\S^2} \int_0^\infty \sum_i r_i\frac{\mathrm{d}N_i}{\mathrm{d}E} A_{\rm eff}(E)\,\mathrm{d}E.
\end{align}
Here $k\in\{\rm ON, OFF\}$ is a label indicating the region on the sky (signal/ON or background/OFF), whereas $\theta_{k,{\rm BG}}$ and $\theta_{k,{\rm DM}}$ are the expected number of events in region $k$ from background processes and DM annihilation, respectively.  These events are photons in the case of DM annihilation, but will be mostly cosmic rays in the case of the background.  The term $\mathrm{d}N_i/\mathrm{d}E$ is the differential photon yield from the $i$th annihilation channel.  We assume an integration time $t_{\rm obs}$ of 200 hours.

The Ring Method, and ON-OFF analyses generally, are designed to consider the difference between the observed rates in the signal and background regions.  If the background rate is expected to be uniform across the entire ring, then after correction for the ratio of sky areas covered by the signal and background regions $\alpha\equiv \Delta\Omega_{\rm ON} / \Delta\Omega_{\rm OFF}$, the expected difference in the observed counts reflects only signal processes 
\begin{align}
\theta_{\rm diff} &\equiv \theta_{\rm ON} - \alpha\theta_{\rm OFF} \nonumber\\
                  &= \theta_{\rm ON, BG} + \theta_{\rm ON,DM} - \alpha\theta_{\rm OFF, BG} - \alpha\theta_{\rm OFF, DM} \nonumber\\
                  &= \theta_{\rm ON, DM} - \alpha\theta_{\rm OFF, DM}. 
\end{align}
In the case of the ring geometry that we adopted for Array B from ref.\
\cite{CTA}, $\Delta\Omega_{\rm ON} = 9.97\times10^{-4}$\,sr,
$\Delta\Omega_{\rm OFF} = 4.05\times10^{-3}\,{\rm sr} \implies \alpha =
0.246$.  Our value of $\Delta\Omega_{\rm ON}$ is $\sim$4\% smaller than stated
in ref.\ \cite{CTA}, but this can likely be explained by the number of
significant figures with which ref.~\cite{CTA} gave their optimized Ring Method parameters.

We model the likelihood of observing a given difference $N_{\rm diff} \equiv N_{\rm ON} - \alpha N_{\rm OFF}$ between the ON-region and scaled OFF-region counts, as the difference of two Poisson processes.  This is known as a Skellam distribution \cite{Skellam}:
\begin{align}
\mathcal{L}_{\rm S}(N_{\rm diff}|\theta_{\rm ON},\alpha\theta_{\rm OFF}) &= e^{-(\theta_{\rm ON}+\alpha\theta_{\rm OFF})}\left(\frac{\theta_{\rm ON}}{\alpha\theta_{\rm OFF}}\right)^\frac{N_{\rm diff}}{2} \nonumber\\ &\times I_{|N_{\rm diff}|}(2\sqrt{\alpha\theta_{\rm ON}\theta_{\rm OFF}}),
\label{skellamlike}
\end{align}
where $I_n$ is the $n$th modified Bessel function of the first kind.  To determine the \textit{expected} limit as we do here, one simply calculates this likelihood assuming $N_{\rm diff} = 0$.  Because the dominant background for CTA comes from misidentified electron events, to obtain $\theta_{k,{\rm BG}}$ we model the expected background flux $\Phi_\mathrm{BG}$ with an electron spectrum $E^3\Phi_\mathrm{BG} = 1.5 \times 10^{-2}$\,GeV$^2$\,cm$^{-2}$\,s$^{-1}$\,sr$^{-1}$, as seen by \Fermi\ \cite{Abdo:2009zk}.  Our final effective likelihood function is the ratio of the signal$+$background likelihood function 
(eq.\ \ref{skellamlike}) to the background-only version
\be
\mathcal{L}_{\rm CTA}(m_\S,\lhs) = \frac{ \mathcal{L}_{\rm S}\left[0|\theta_{\rm ON}(m_\S,\lhs),\alpha\theta_{\rm OFF}(m_\S,\lhs)\right] }
                                       { \mathcal{L}_{\rm S}(0|\theta_{\rm ON,BG},\alpha\theta_{\rm OFF,BG}) }. 
\label{ctalike}
\ee
In deriving expected limits we know the best-fit likelihood to occur where the signal contribution is zero, so 
eq.\ (\ref{ctalike}) has a maximum $\mathcal{L}_{\rm CTA}=1$ at $\langle
\sigma v_{\rm rel} \rangle_0 = 0$.  The Skellam distribution is already almost a Gaussian, so by the Central Limit Theorem the ratio 
eq.\ (\ref{ctalike}) is very close to Gaussian.  We can therefore safely consider this likelihood ratio to be $\chi^2$-distributed with one degree of freedom, and derive confidence limits accordingly.

\bibliographystyle{apsrev}

\end{document}